\documentclass[11pt,a4paper]{article} 
\pdfoutput=1

\usepackage{amsmath,amsfonts,amsbsy,amssymb,array,accents}
\usepackage{enumerate,array,latexsym,graphicx,mathrsfs,verbatim,psfrag}
\usepackage{bm} 
\usepackage[normalem]{ulem}
\usepackage{booktabs} 
\usepackage[usenames]{color}

\usepackage[english]{babel}
\usepackage{fancybox}

\usepackage[nosort]{cite}
\usepackage{chngpage} 

\usepackage[utf8]{inputenc}

\usepackage{enumitem}
\setlist{itemsep=0pt}

\usepackage[colorlinks=true,      linkcolor=darkblue,      urlcolor=darkblue,      
            filecolor=darkblue,      citecolor=darkblue,       pdfstartview=FitH,     
						pdfpagemode=UseNone,      bookmarksopen=true]{hyperref}  
\usepackage[all]{hypcap}     

\def\eq#1{(\ref{#1})}

\usepackage{mathtools}
\DeclarePairedDelimiter\bra{\langle}{\rvert}
\DeclarePairedDelimiter\ket{\lvert}{\rangle}
\DeclarePairedDelimiterX\braket[2]{\langle}{\rangle}{#1 \delimsize\vert #2}

\newcommand{\captionfonts}{\small}
\makeatletter  
\long\def\@makecaption#1#2{%
  \vskip\abovecaptionskip
  \sbox\@tempboxa{{\captionfonts #1: #2}}%
 \ifdim \wd\@tempboxa >\hsize
    {\captionfonts #1: #2\par}
  \else
    \hbox to\hsize{\hfil\box\@tempboxa\hfil}%
  \fi
  \vskip\belowcaptionskip}
\makeatother   

\DeclareMathSymbol{\medhatsym}{\mathord}{largesymbols}{"62} 

\DeclareMathSymbol{\medtildesym}{\mathord}{largesymbols}{"65}
\newcommand\lowermedtildesym{
  \text{\smash{\raisebox{-1.2ex}{%
    $\medtildesym$}}}}
\newcommand\medtilde[1]{
  \mathchoice
    {\accentset{\displaystyle\lowermedtildesym}{#1}}
    {\accentset{\textstyle\lowermedtildesym}{#1}}
    {\accentset{\scriptstyle\lowermedtildesym}{#1}}
    {\accentset{\scriptscriptstyle\lowermedtildesym}{#1}}
}

\mathchardef\mhyphen="2D

\def\({\left(}
\def\){\right)}
\def\[{\left[}
\def\]{\right]}

\def\barray{\begin{array}}
\def\earray{\end{array}}
\def\be{\begin{equation}}
\def\ee{\end{equation}}
\def\bea{\begin{eqnarray}}
\def\eea{\end{eqnarray}}
\def\bal{\begin{align}}
\def\eal{\end{align}}

\numberwithin{equation}{section} %

\makeatletter
\g@addto@macro\bfseries{\boldmath}
\makeatother

\definecolor{cardinal}{rgb}{0.6,0,0}
\definecolor{darkgreen}{rgb}{0,0.4,0}
\definecolor{purple}{rgb}{0.5, 0, 0.5}
\definecolor{golden}{rgb}{0.92, 0.7, 0}
\definecolor{midnight}{rgb}{0, 0, 0.5}
\definecolor{darkblue}{rgb}{0, 0, 0.8}

\def\cB{{\cal B}}

\def\cF{{\cal F}}

\def\cM{{\cal M}}

\def\cP{{\cal P}}

\def\sst#1{\scriptscriptstyle{#1}}

\newcommand{\NS}{\ensuremath{\mathrm{\sst{NS}}}}

\newcommand{\betab}{\ensuremath{\bm{\beta}}}
\newcommand{\Aform}{\ensuremath{\mathbf{A}}}
\newcommand{\Bform}{\ensuremath{\mathbf{B}}}

\def\coeff#1#2{{\textstyle \frac{#1}{#2}}}
	
\begin{document}

\begin{flushright}
%
%
\end{flushright}

\vspace{19mm}

\begin{center}

{\huge \bf{AdS$_3$ Holography at Dimension Two}}

\vspace{22mm}

{\large
\textsc{Stefano Giusto$^{1,2}$, ~Sami Rawash$^3$, ~David Turton$^3$}
}

\vspace{15mm}

${}^{1}$Dipartimento di Fisica ed Astronomia ``Galilei Galilei'', Universit\`a di Padova,\\
Via Marzolo 8, 35131 Padova, Italy\\

\vspace{5mm}

${}^{2}$I.N.F.N. Sezione di Padova, Via Marzolo 8, 35131 Padova, Italy\\

\vspace{5mm}

${}^{3}$Mathematical Sciences and STAG Research Centre, University of Southampton,\\
Highfield, Southampton SO17 1BJ, United Kingdom

\vspace{7mm}

{\footnotesize\upshape\ttfamily  stefano.giusto @ unipd.it, ~s.rawash @ soton.ac.uk, ~d.j.turton @ soton.ac.uk } \\

\vspace{20mm}

\textsc{Abstract}
\vspace{7mm}
\begin{adjustwidth}{15mm}{15mm} 
\noindent
Holography can provide a microscopic interpretation of a gravitational solution as corresponding to a particular CFT state:  
the asymptotic expansion in gravity encodes the expectation values of operators in the dual CFT state. 
Such a correspondence is particularly valuable in black hole physics. 
We study supersymmetric D1-D5-P black holes, for which recently constructed microstate solutions known as ``superstrata'' provide strong motivation to derive the explicit D1-D5 holographic dictionary for CFT operators of total dimension two.
In this work we derive the explicit map between one-point functions of scalar chiral primaries of dimension (1,1) and the asymptotic expansions of families of asymptotically AdS$_3\times S^3 \times \mathcal{M}$ supergravity solutions, with $\mathcal{M}$ either T$^4$ or K3.
We include all possible mixings between single-trace and multi-trace operators. 
We perform several tests of the holographic map, including new precision holographic tests of superstrata, that provide strong supporting evidence for the proposed dual CFT states.
\end{adjustwidth}

\end{center}

\thispagestyle{empty}

\newpage


%
%


\baselineskip=15.1pt
\parskip=3pt


\section{Introduction}
\label{sec:intro}
Holography has been instrumental in enlightening the microscopic properties of black holes. Even before holographic duality was formulated in its mature form \cite{Maldacena:1997re, Witten:1998qj,Gubser:1998bc}, the first microscopic derivation \cite{Strominger:1996sh} of the entropy of a black hole with a finite-size classical horizon -- the supersymmetric D1-D5-P black hole -- was based on the counting of states in a brane worldvolume conformal field theory (CFT). Making progress towards a solution of the black hole information paradox~\cite{Hawking:1976ra} requires going beyond the black hole microstate counting problem and understanding how the properties of individual black hole microstates manifest themselves in the spacetime/bulk description, rather than in the dual CFT description. While much of the recent literature emphasizes the consequences for the gravitational description of some universal properties of the CFT, most notably its Virasoro algebra (see for instance \cite{Fitzpatrick:2014vua,Hartman:2014oaa,Anous:2016kss,Fitzpatrick:2016ive}), it is natural to expect that the guide provided by an explicit string-theoretical model of a black hole, such as the D1-D5-P black hole, might be crucial to elucidate the fine-grained structure of the microstates, which is ultimately responsible for the unitarity of black hole evaporation. 

In this article we will exploit the power of holography in elucidating the properties of black hole microstates, focusing on the D1-D5-P black hole. The two sides of the holographic duality involve the decoupling region of the black hole geometry, which is asymptotically AdS$_3\times S^3\times \mathcal{M}$, with $\mathcal{M}$ either $T^4$ or $K3$, and a 2D CFT with (4,4) supersymmetry, known as the D1-D5 CFT \cite{David:2002wn,Avery:2010qw}. We will work in the best controlled limit of the holographic duality, and thus ignore $1/N$ and $\alpha'$ corrections. As usual in gauge/gravity duality, classical supergravity is dual to a strongly coupled point in the CFT moduli space, while field-theory calculations are tractable around a free locus where the D1-D5 CFT reduces to a supersymmetric orbifold sigma-model. For this reason, the possibility of building a precise map between CFT states and classical supergravity configurations rests on the existence of moduli-independent quantities, which can be defined only for supersymmetric states; however at least at the qualitative level, the insights coming from this holographic analysis are expected to be useful also for non-supersymmetric black holes. Supersymmetric microstates of the D1-D5-P black hole are dual to ``heavy'' CFT operators that preserve 1/8 of the 32 supercharges of type IIB supergravity and with conformal dimensions that scale as the central charge $c=6 N$ in the large $N$ limit, with $N=n_1 n_5$ given by the product of the integer numbers of D1 and D5 branes. Working at large $N$ guarantees that the mass of the states is large in Planck units. In principle one can completely characterize a heavy CFT state $|H\rangle$ by giving the expectation values of all operators $O_i$ in the state, i.e.
\begin{equation}
\langle H | O_i | H \rangle\,.
\end{equation}
These expectation values are in general non-trivial functions of the CFT moduli, however a non-renormalization theorem proved in \cite{Baggio:2012rr} guarantees that when $O_i$ is a chiral primary operator and $|H\rangle$ a 1/4 or 1/8 BPS state, the expectation values are protected quantities that do not depend on the moduli. Hence such one-point functions computed at the free orbifold point can be matched with supergravity where, according to the AdS/CFT dictionary, they are encoded in the deviations of the geometry from pure AdS, in an expansion around the asymptotic boundary. 

The holographic point of view then implies that, on the gravity side, individual microstates can in principle be distinguished from each other and from the classical black hole geometry by the asymptotic fall-off of the fields in the decoupling region of the geometry (including in principle the non-supergravity fields of string theory). This reasoning provides part of the conceptual basis for the fuzzball program \cite{Lunin:2001fv,Lunin:2001jy,Lunin:2002iz,Bena:2007kg,Skenderis:2008qn,Balasubramanian:2008da,Bena:2013dka}, which associates regular bulk string theory solutions (that are not necessarily well-described in supergravity) to the heavy CFT states $|H\rangle$. Some comments are in order. Firstly, while large sets of states may be distinguished solely through the expectation values of chiral primaries, of course many states are not distinguished by such expectation values. Allowing $O$ to be any CFT operator requires going beyond the supergravity approximation. The stringy description of black hole microstates beyond supergravity is an interesting open avenue, with recent progress~\cite{Martinec:2017ztd,Martinec:2018nco}, however developing a precise holographic dictionary in this more general context appears to be a formidable task. Secondly, for a typical microstates $|H\rangle$ in a given ensemble, the deviation of the expectation values $\langle H | O_i | H \rangle$ from the ensemble average values are expected to be exponentially suppressed in the large $N$ limit (see e.g.~\cite{Lashkari:2014pna}), at least for a simple enough $O_i$. This implies that the states that admit a simple semiclassical gravitational description are necessarily somewhat atypical. Nevertheless, by approaching typical microstates through limits of progressively less atypical microstates which are amenable to study, one hopes to gain valuable insights about the structure of typical states. For a recent discussion of related points, see~\cite{Mathur:2018tib}.

The holographic analysis of black hole microstates based on the expectation values $\langle H | O_i | H \rangle$ was pioneered in \cite{Skenderis:2006ah,Kanitscheider:2006zf,Kanitscheider:2007wq,Taylor:2007hs}, building on the results of~\cite{Skenderis:2006uy,Taylor:2005db}. Those works laid out a general formalism for precision holography, and the holographic dictionary was made explicit when $|H\rangle$ is a 1/4 BPS state, carrying D1 and D5 but no P charge, and when $O_i$ is a set of  chiral primary operators with  dimension less than or equal to two. The holographic dictionary was further developed and extended to a particular set of 1/8 BPS states in \cite{Giusto:2015dfa}, however only for chiral primaries of dimension one. 

Holographic studies involving expectation values of operators of total dimension two present some interesting complications, both technical and conceptual~\cite{Taylor:2007hs}. On one hand, for some operators of dimension two, the correspondence between CFT operators and supergravity fields cannot be uniquely fixed solely on the basis of the quantum numbers, and in the holographic dictionary operators $O_i$ with the same quantum numbers may mix. On the other hand, while in the previous discussion we implicitly assumed that $O_i$ is a single-particle (single-trace) operator, one can also form dimension-two operators by taking the product of two single-trace operators of dimension one; these double-trace operators can also mix with single-trace operators and, while the multi-trace contributions are suppressed by powers of $1/N$ in generic correlators, they can contribute at leading order to `extremal' correlators (a correlator is extremal if the dimension of one operator is equal to the sum of the dimensions of the others.). Both these issues were addressed in \cite{Taylor:2007hs}, where a precise form of the mixing between single-trace operators was derived, and more qualitative results regarding the mixing with double-trace operators were proposed.

In this article we construct a fully explicit holographic dictionary for operators of dimension $(h,\bar{h}) = (1,1)$, that can be used as a quantitative tool to perform new precision tests of the whole class of 1/4 and 1/8 BPS microstates currently known: these include the D1-D5 geometries constructed in \cite{Lunin:2001jy,Lunin:2002iz,Kanitscheider:2007wq} and also more recently constructed three-charge supergravity solutions including those known as ``superstrata''~\cite{Bena:2015bea,Bena:2016agb,Bena:2016ypk,Bena:2017geu,Bena:2017upb,Bena:2017xbt,Bena:2018bbd,Bakhshaei:2018vux,Bena:2018mpb,Ceplak:2018pws,Heidmann:2019zws} and related solutions~\cite{Mathur:2011gz,Mathur:2012tj,Lunin:2012gp,Giusto:2013bda}. As an application, we perform new holographic tests of the proposed dual CFT description of a set of superstrata.

Our construction proceeds by first carefully defining the normalizations of both the CFT operators and the coefficients in the asymptotic expansion of the supergravity solutions, and then by fixing the numerical coefficients defining the holographic dictionary by matching the CFT predictions with some reference D1-D5 geometries, whose identification with CFT states is already well-established. This determines uniquely both the mixing amongst the single-trace operators, in precise agreement with the results of \cite{Taylor:2007hs}, and the mixing between single and double-trace operators. We then test the holographic map against other microstates, including both D1-D5 and D1-D5-P states; a stringent requirement comes from the invariance under the R-symmetry group, which implies that the coefficients defining the holographic map should be the same for all the operators in the same R-charge multiplet. We will see that the consistency of the dictionary often works in a non-trivial fashion, thus providing strong supporting evidence of the proposed identification between the CFT states and the dual supergravity solutions. 

While this paper was in the final stages of preparation for publication, we received \cite{Tormo:2019yus} that contains some related calculations of one-point functions in the D1-D5 orbifold CFT.

The remainder of this paper is organized as follows.
We review the correspondence between 1/4-BPS coherent states of the orbifold CFT and the family of D1-D5 supergravity solutions in Section~\ref{sec:states}. The holographic map for chiral primary operators (CPOs) of dimension 1 is summarized in Section~\ref{sec:dim1}; this is mostly a recollection of previous results \cite{Skenderis:2006ah,Kanitscheider:2006zf,Kanitscheider:2007wq,Giusto:2015dfa}, however we clarify some minus signs that are needed to make the dictionary for the $SU(2)_L\times SU(2)_R$ R-currents consistent. In Section~\ref{sec:dim2} we describe all the CPOs of dimension $(1,1)$, including single and double-trace operators, and we first work out the holographic dictionary for the simpler subsector of operators, which does not involve mixing between different single-traces. The more complicated subsector is analyzed in Section~\ref{sec:sigmaomega}, in which we fix in turn each of the coefficients defining the holographic dictionary, and then make some non-trivial tests on 1/4 BPS states. In Section~\ref{sec:superstrata} we apply our results to perform new precision holographic tests of D1-D5-P superstrata. We comment on the significance of our results for the fuzzball program, and on possible future developments, in the Discussion in Section \ref{sec:disc}. In the Appendices we record our conventions for S$^3$ spherical harmonics, the derivation of some CFT correlators involving twist fields, and some details of the general class of D1-D5-P supersymmetric supergravity solutions that are invariant on the internal manifold $\cM$.

\section{Holography for D1-D5 black hole microstates}
\label{sec:states}

In this section we give a brief review of holography for D1-D5 black hole microstates, with the main purpose of setting up notation that is needed in the rest of the paper. 

The dual gravitational description of the Ramond-Ramond (RR) ground states of the D1-D5 CFT is well known~\cite{Lunin:2001fv,Lunin:2001jy,Lunin:2002bj,Lunin:2002iz,Skenderis:2006ah,Kanitscheider:2007wq}. There is a family of supergravity solutions that can be associated with coherent RR ground states of the D1-D5 CFT, in the sense that protected correlators involving such states agree, as discussed in the Introduction.

The states of the D1-D5 CFT have a simple description at the free orbifold locus in moduli space, where the CFT is the $(4,4)$ sigma-model with target space $\mathcal{M}^N/S_N$, with $\mathcal{M}$ either T$^4$ or K3 (recall $N=n_1 n_5$). A review of the orbifold CFT can be found for example in \cite{Avery:2010qw}. In this article we will use the notation and the conventions of \cite{Giusto:2015dfa,Bena:2017xbt}. A generic state of the orbifold CFT is described by a collection of ``strands" involving spin-twist operators; the ground state of each strand is characterized by a spin $s$ and a winding number $k$ and is denoted by $\ket{s}_k$. In this article we will consider bosonic ground states, and excitations thereof, that are insensitive to the structure of the internal manifold $\mathcal{M}$, so that our results apply when $\mathcal{M}$ is either T$^4$ or K3 (the generalization to more general states is straightforward). For this class of ground states, there are five possible spin configurations: $s=(0,0), (\pm,\pm)$, where $(j,\bar j)$ denotes a state with $SU(2)_L$ charge $j$ and $SU(2)_R$ charge $\bar j$; $SU(2)_L\times SU(2)_R$ is the R-symmetry of the $(4,4)$ theory, which corresponds on the gravity side to rotations in the four spatial directions. A RR ground state with $N_k^{(s)}$ strands of type $\ket{s}_k$ is denoted by
\begin{equation}\label{eq:RRstate}
\psi_{\{ N_k^{(s)}\}} \equiv \prod_{k,s} (\ket{s}_k)^{N_k^{(s)}}\,,
\end{equation}
and is an allowed state if the total winding number sums up to $N$:
\begin{equation}\label{eq:windingconstraint}
\sum_{k,s} k N_k^{(s)} = N\,.
\end{equation}
It will be convenient to work with non-normalized states; for later use we record the norm of the states \eqref{eq:RRstate}, which was derived in \cite{Giusto:2015dfa}:
\begin{equation}\label{eq:norm}
 \left |\psi_{\{ N_k^{(s)}\}}\right |^2 = \frac{N!}{\prod_{k,s} N_k^{(s)}!\, k^{N_k^{(s)}}}\,.
\end{equation}
States of the form \eqref{eq:RRstate} are eigenstates of the $SU(2)_L\times SU(2)_R$ currents $(J^3, \tilde J^3)$; we are interested in coherent states that are linear combinations of R-symmetry eigenstates labeled by complex coefficients $A_k^{(s)}$,
\begin{equation}\label{eq:coherent}
\psi(\{ A_k^{(s)} \}) \equiv {\sum_{\{ N_k^{(s)} \}}}^{\!\!\!\prime} \prod_{k,s} (A_k^{(s)} \ket{s}_k)^{N_k^{(s)}}\,,
\end{equation}
where the sum ${\sum_{\{ N_k^{(s)} \}}}^{\!\!\!\prime}$ is restricted by the constraint \eqref{eq:windingconstraint}. The states that admit a good supergravity description are those for which this sum is peaked over large values of $N_k^{(s)}$: as shown in \cite{Giusto:2015dfa}, in this semiclassical limit the parameters $A_k^{(s)}$ determine the average numbers $\overline{N}_k^{(s)}$ of strands of type $\ket{s}_k$, via
\begin{equation}\label{eq:average}
k \,\overline{N}_k^{(s)} = |A_k^{(s)}|^2\,.
\end{equation}
The constraint \eqref{eq:windingconstraint} then implies
\begin{equation}\label{eq:windingconstraintbis}
\sum_{k,s}|A_k^{(s)}|^2=N\,.
\end{equation}

The supergravity solutions describing coherent bound states of large numbers of D1 and D5 branes are well-known and are given in terms of a profile function $g_i(v')$ in $\mathbb{R}^8$~\cite{Lunin:2001jy,Lunin:2002iz,Kanitscheider:2007wq}. For configurations invariant on the internal manifold $\cM$, the profile function takes values in $\mathbb{R}^5$:
\begin{equation}\label{eq:FPcurve}
\begin{aligned}
g_1(v')+i g_2 (v') &=\sum_{k>0}\left( \frac{\bar a_k^{(++)}}{k} e^{\frac{2\pi i k}{L} v'} + \frac{{a}_k^{(--)}}{k} e^{-\frac{2\pi i k}{L} v'} \right),\\
g_3(v')+i g_4 (v') &= \sum_{k>0}\left( \frac{\bar a_k^{(+-)}}{k} e^{\frac{2\pi i k}{L} v'} - \frac{{a}_k^{(-+)}}{k} e^{-\frac{2\pi i k}{L} v'}\right) ,\\
g_5(v') &= - \mathrm{Im} \left(\sum_{k>0} \frac{\bar a_k^{(00)}}{k} e^{\frac{2\pi i k}{L} v'} \right).
\end{aligned}
\end{equation}

The map between the CFT states in~\eq{eq:coherent} and the supergravity solutions parameterized by the profile $g_i(v')$ which will be described in more detail below, is given by relating the Fourier modes $a^{(s)}_k$ to the coherent state parameters $A^{(s)}_k$, via\footnote{We note that in Eq.\;\eq{eq:FPcurve}, the minus sign in front of $a_k^{(-+)}$ and the complex conjugations differ from those given in~\cite{Giusto:2015dfa}. We will see in due course that these details in Eq.\;\eq{eq:FPcurve} are needed for consistency of the holographic map \eq{eq:Aamap} and the rest of our conventions.}
\begin{equation}\label{eq:Aamap}
A_k^{(\pm\pm)} = R\sqrt{\frac{N}{Q_1 Q_5}} a_k^{(\pm\pm)}~,\qquad A_k^{(00)}=R\sqrt{\frac{N}{2\,Q_1 Q_5}} a_k^{(00)}\;.
\end{equation}

The curve $g_i(v')$ arises because the D1-D5 system is U-dual to a fundamental string (F1) carrying momentum (P): in the F1-P duality frame, the curve \eqref{eq:FPcurve} represents the oscillation profile of the string in the five transverse directions that are U-dual to D1-D5 states invariant on $\mathcal{M}$.
The D1-D5 supergravity solution associated with a curve $g_i(v')$ is as follows. The 6D Einstein metric of this solution is given by
\begin{equation}\label{eq:metric6D}
ds_6^2 \;=\; - \frac{2}{\sqrt{\mathcal{P}}}(dv+ \betab )\left(du+\omega + \frac{\mathcal{F}}{2}(dv + \betab )\right) + \sqrt{\mathcal{P}} ds_4^2 \,,
\end{equation}
with 
\begin{equation}
\mathcal{P}\,=\, Z_1 Z_2 - Z_4^2\,.
\end{equation}
The 4D metric $ds_4^2$ describes the four spatial non-compact directions $x_i$, and, for all the solutions considered in this article, is the flat $\mathbb{R}^4$ metric
\begin{equation}
ds_4^2 \,=\, dx_i dx_i\,.
\end{equation} 
The $u$ and $v$ coordinates parametrize time $t$ and the $S^1$ direction $y$, which we take to have radius $R_y$:
\begin{equation} \label{eq:ty}
u \,\equiv\, \frac{t-y}{\sqrt{2}}\quad,\quad v \,\equiv\, \frac{t+y}{\sqrt{2}}\,.
\end{equation}
The D1 and D5 charges of the solution are given by 
\begin{equation}\label{eq:Q1Q5}
Q_1 \,=\, \frac{(2\pi)^4 \,n_1 \,g_s\, \alpha'^4}{V_4}\,,\qquad Q_5 \,=\, n_5 \,g_s \,\alpha'\,,
\end{equation}
where $g_s$ is the string coupling, and $V_4$ is the coordinate volume of $\mathcal{M}$. The periodicity $L$ of the curve $g_i(v')$ is $L=2\pi Q_5 /R_y$. The solution is specified by the scalar functions $Z_1$, $Z_2$, $Z_4$ and $\mathcal{F}$ and by the 1-forms with legs along $\mathbb{R}^4$, $\betab$ and $\omega$. The solutions dual to RR ground states have $\mathcal{F}=0$ and all the other scalars and 1-forms are only functions of $x_i$, specified by the curve $g_i(v')$ as follows:
\begin{equation}\label{eq:metricdata}
\begin{aligned}
&Z_1 \;=\; \frac{Q_5}{L}\int_0^L dv' \,\frac{|\dot{g}_i(v')|^2+ |\dot{g}_5(v')|^2}{|x_i - g_i(v')|^2}~,\qquad Z_2 \;=\; \frac{Q_5}{L}\int_0^L dv' \,\frac{1}{|x_i - g_i(v')|^2}\,,\\
&Z_4 \;=\; - \frac{Q_5}{L}\int_0^L dv' \,\frac{\dot{g}_5(v')}{|x_i - g_i(v')|^2}~,\qquad\quad~~ \Aform\;=\; -  \frac{Q_5}{L}\int_0^L dv' \,\frac{\dot{g}_j(v') dx_j}{|x_i - g_i(v')|^2}\,,\\
&d\Bform\;=\; - *_4 d\Aform~,\qquad~ \betab = \frac{-\Aform+\Bform}{\sqrt{2}}~,\qquad~ \omega\;=\; -\frac{\Aform+\Bform}{\sqrt{2}}\,,
\end{aligned}
\end{equation}
where the dot indicates the derivative with respect to $v'$ and $*_4$ is the Hodge dual with respect to the flat metric $ds_4^2$. Besides the 6D metric $ds_6^2$ in \eqref{eq:metric6D}, the solution contains all other NSNS and RR fields of type IIB supergravity: their form is entirely specified by the curve $g_i(v')$ through the above functions, and is recorded for completeness in Eq.\;\eq{ansatzSummary}.

In summary, the geometry dual to the RR ground state \eqref{eq:coherent} is completely specified by the curve $g_i(v')$ \eqref{eq:FPcurve}, through Eqs.\;\eqref{eq:metric6D}--\eqref{eq:metricdata}. 
Given the identification between gravity and CFT parameters in Eq.\;\eq{eq:Aamap}, the CFT constraint \eqref{eq:windingconstraintbis} becomes
\begin{equation}\label{eq:regularityconstraint}
\sum_{k>0}\left(|a_k^{(++)}|^2 + |a_k^{(--)}|^2+|a_k^{(+-)}|^2+|a_k^{(-+)}|^2+\frac{1}{2}|a_k^{(00)}|^2 \right)\;=\;\frac{Q_1 Q_5}{R_y^2}\,,
\end{equation}
which, on the gravity side, is the regularity condition for the solution \eqref{eq:metric6D}--\eqref{eq:metricdata}.

The holographic map can also be extended to a subset of the BPS states carrying D1, D5 and momentum (P) charge, which in CFT terms are states with $L_0>\tilde L_0= \frac{c}{24}$. 
There is not yet a general understanding of the full class of D1-D5-P states, however there has been much recent progress in constructing large families of explicit solutions known as ``superstrata''~\cite{Bena:2015bea,Bena:2016agb,Bena:2016ypk,Bena:2017geu,Bena:2017upb,Bena:2017xbt,Bena:2018bbd,Bakhshaei:2018vux,Bena:2018mpb,Ceplak:2018pws,Heidmann:2019zws}. There is an explicit proposal for the dual CFT states of these solutions~\cite{Bena:2015bea,Bena:2016agb,Bena:2016ypk,Bena:2017xbt,Ceplak:2018pws}.
This family of solutions, and the proposed map to states of the orbifold CFT, will be reviewed in Section~\ref{sec:superstrata}. Their 6D metric can still be written in the form \eqref{eq:metric6D} with a flat $ds_4^2$, but now $\mathcal{F}\not=0$ and the scalars and 1-forms specifying the solution are functions of $v$ as well as $x_i$. Given the similarities of the supergravity description of this class of D1-D5-P states with the D1-D5 states, one can formulate a unified recipe to extract expectation values of operators of dimension one and two from the geometry. We proceed to do this in the next three sections.

\section{Expectation values of operators of dimension one}
\label{sec:dim1}

In this section we review the holographic map for expectation values of operators of dimension one, making precise some details that will be important in the following sections.

We start by setting up some notation for the field content of the D1-D5 orbifold CFT.
We label the $N$ copies of the CFT on $\mathcal{M}$ by the index $r=1,\ldots,N$. The orbifold CFT has R-symmetry group $SU(2)_L\times SU(2)_R$, whose spinorial indices we denote by $\alpha,\dot \alpha= \pm$, and there is also an $SU(2)_1\times SU(2)_2$ group of rotations on the tangent space of $\mathcal{M}$ that is useful for labelling operators, whose spinorial indices we denote by $A, \dot A= 1,2$.
On each copy of the CFT, the fundamental fields are four bosons $X^{A\dot A}_{(r)}$, and four left-moving plus four right-moving fermions $\psi^{\alpha\dot A}_{(r)}$, $\tilde \psi^{\dot\alpha\dot A}_{(r)}$. 

The theory also contains spin-twist operators, that change the boundary conditions of the fields, and that are labelled by permutations of $S_N$. For example, the `bare' twist operator $\sigma_{(rs)}$ joins or splits the copies $r$ and $s$. When acting on untwisted strands in their respective NS vacuum state, $\sigma_{(rs)}$ creates the state that is the lowest state on a twist-two strand, which is the NS vacuum of the two-fold covering space. A brief review of covering space methods and a more general definition of spin-twist operators is given in Appendix \ref{sec:twistfieldcorr}. We also have left and right-moving spin-fields $S^{\alpha}$, $\bar{S}^{\dot\alpha}$ in each twisted sector, that map NS ground states to R ground states.

Though this description in terms of free fields ceases in general to be useful away from the orbifold point, there are physical quantities that are guaranteed to be independent of the moduli, and hence can be quantitatively described by the free orbifold CFT. In particular in this paper we will focus on the expectation values of chiral primary operators (CPOs) and their (global) $SU(2)_L\times SU(2)_R$ descendents in states preserving eight or four supercharges \cite{Baggio:2012rr}: the first class of states are the RR ground states described in the previous section and the states in the second class include the D1-D5-P states that will be considered in Section~\ref{sec:superstrata}. Note that in both classes, the states are ``heavy", in the sense that their left and right dimensions $h$ and $\bar h$ are of order of the CFT central charge $c=6 N$: one has $h=\bar h=c/24$ for the D1-D5 states and $h>\bar h=c/24$ for the D1-D5-P states. The CPOs we will consider are instead ``light", having $h,\bar h$ of order $c^0$. In particular we will restrict to CPOs with $h+\bar h\le 2$. The purpose of the next two sections is to formulate and test a recipe to compute the expectation values of light CPOs in heavy states from the asymptotic expansion of the geometries dual to the heavy states.

Expectation values of CPOs with total dimension $\Delta=h+\bar h=1$ have already been considered in \cite{Kanitscheider:2006zf,Kanitscheider:2007wq,Giusto:2015dfa}. 
The only operators with $h=\pm j=1$, $\bar h=\bar j=0$ are the $SU(2)_L$ generators $J^\pm$:
\begin{equation}
J^\pm = \sum_r J^\pm_{(r)}=\pm \sum_r \psi^{\pm 1}_{(r)} \psi^{\pm 2}_{(r)}\,; 
\end{equation}
analogously one has the $SU(2)_R$ generators $\tilde J^\pm$, with $h=j=0$, $\bar h=\pm \bar j=1$:
\begin{equation} \label{eq:jtildepm}
\tilde J^\pm  = \sum_r \tilde J^\pm_{(r)}=\pm \sum_r \tilde \psi^{\pm 1}_{(r)} \tilde \psi^{\pm 2}_{(r)}\,.
\end{equation}
We define $J^3$ to be normalized according to the standard commutation relation $[J^+,J^-]=2J^3$ and such that the eigenvalue of $J^3$ on the RR ground state $\ket{\pm+}$ is $\pm 1/2$; similarly for ${\tilde J}^3$; see Appendix \ref{sec:twistfieldcorr} for more details. We normalize the corresponding vector spherical harmonics in the same way, see Appendix \ref{sec:sphericalharm} for details. Note that this convention means that the normalized affine descendant of $J^+$ is $\frac{1}{\sqrt{2}}[J^-,J^+]=-\sqrt{2}J^3$, which means that some factors of $\sqrt{2}$ will show up in equations such as \eq{eq:mapdim1}.

Next we have the operators with $h=j=\bar h=\bar j=1/2$. The first of these is the twist-two operator
\begin{equation}\label{eq:Sigma2}
\Sigma_2^{++} \;=\; \sum_{r<s} \sigma_{(rs)}^{++}\,, \qquad\quad \sigma_{(rs)}^{++} \;=\; S^+_{(rs)} \bar{S}^+_{(rs)} \sigma_{(rs)}
\end{equation}
where the operator $\sigma_{(rs)}$ is the `bare' twist operator that joins or splits the copies $r$ and $s$, and $S^+_{(rs)}$, $\bar{S}^+_{(rs)}$ are spin fields. When acting on untwisted strands in the NS vacuum state, $\sigma_{(rs)}^{++}$ creates the twisted RR vacuum state $\ket{++}_{2\,}$.

The second chiral primary with $h=j=\bar h=\bar j=1/2$ is the untwisted operator
\begin{equation}\label{eq:opO}
O^{++} \;=\;\sum_r O^{++}_{(r)}\;=\; \sum_{r} \frac{1}{\sqrt{2}} \epsilon_{\dot A \dot B} \,\psi^{+\dot A}_{(r)} \tilde \psi^{+\dot B}_{(r)}\,.
\end{equation}
More generally, one has operators like in \eqref{eq:opO} for each of the $h^{1,1}(\mathcal{M})$ elements of the $(1,1)$ cohomology of $\mathcal{M}$: we focus on the unique $SU(2)_1\times SU(2)_2$ operator $O^{++}$ because it is the only one that has non-trivial expectation values on the $\mathcal{M}$-invariant class of states introduced in Section~\ref{sec:states}.

For any CPO one also has the whole multiplet of (global) $SU(2)_L\times SU(2)_R$ descendants, obtained in the usual way by acting on the CPO with $J^-_0$ and/or $\tilde J^-_0$. We denote the generic elements of the multiplet
by $J^a$, $\tilde J^a$, with $a=+,3,-$, and $O^{\alpha,\dot \alpha}$, $\Sigma^{\alpha,\dot \alpha}$, with $\alpha,\dot\alpha = \pm$. 
For later use we record our convention that $O^{--}= (O^{++})^\dagger$, whereupon consistency
 with the $SU(2)_L\times SU(2)_R$ algebra implies that    $O^{-+}= -(O^{+-})^\dagger$,  since $(O^{+-})^\dagger = ([\tilde J_0^-,O^{++}])^\dagger = -[\tilde J_0^+,O^{--}]=-O^{-+}$. Analogous expressions hold for $\Sigma^{\alpha,\dot \alpha}$.

The expectation values of the CPOs and their descendants in a heavy state are encoded in the asymptotic expansion of the dual geometry near their AdS$_3\times S^3$ boundary. Roughly speaking, given a radial coordinate $r$, operators of increasing dimension correspond to terms of higher order in $1/r$. The precise map involves identifying gauge-invariant quantities~\cite{Skenderis:2006uy,Kanitscheider:2006zf}; having done so, in practice it is convenient to choose a particular gauge in which to work. Though there is, in general, no canonical choice for $r$, for the class of  geometries of the form \eqref{eq:metric6D} with a {\it flat} $ds^2_4$ one can canonically identify $r$ with the radial coordinate of $\mathbb{R}^4$ in standard polar coordinates:
\begin{equation}\label{eq:polarcoord}
ds^2_4 = dr^2 + r^2 (d\theta^2 + \sin^2\theta \,d\phi^2 + \cos^2\theta\, d\psi^2)\,. 
\end{equation}
Similarly we can use the $\theta,\phi,\psi$ coordinates to define spherical harmonics on S$^3$. This leaves us with the only ambiguity of choosing the origin of polar coordinates, which will be fixed shortly. One can then define the following asymptotic expansion \cite{Kanitscheider:2006zf,Kanitscheider:2007wq}:
\begin{equation}\label{eq:geometryexpansion}
\begin{aligned}
Z_1 &=\frac{Q_1}{r^2} \left(1+ \sum_{k=1}^\infty \sum_{m_k,\bar m_k=-k/2}^{k/2} f^1_{k\,(m_k,\bar m_k)} \,\frac{Y_k^{m_k,\bar m_k}}{r^k}\right)\,,\\
Z_2&=\frac{Q_5}{r^2} \left(1+ \sum_{k=1}^\infty \sum_{m_k,\bar m_k=-k/2}^{k/2} f^5_{k\,(m_k,\bar m_k)} \,\frac{Y_k^{m_k,\bar m_k}}{r^k}\right)\,,\\
Z_4&=\frac{\sqrt{Q_1Q_5}}{r^2} \left(\sum_{k=1}^\infty \sum_{m_k,\bar m_k=-k/2}^{k/2} \mathcal{A}_{k\,(m_k,\bar m_k)} \,\frac{Y_k^{m_k,\bar m_k}}{r^k}\right)\,,\\
\Aform&= \frac{\sqrt{Q_1 Q_5}}{r^2}\, \sum_{a=1}^3 (a_{a+} Y_1^{a +} + a_{a-} Y_1^{a -})+O(r^{-3})\,,\qquad \mathcal{F} = -\frac{2 Q_p}{r^2} + O(r^{-3})\,,\!\!\!\!\!\!\!\!
\end{aligned}
\end{equation}
where $Y_k^{m_k,\bar m_k}$ are S$^3$ scalar harmonics of degree $k$ and $Y_1^{a \pm}$ are vector harmonics of degree 1; we list our definitions and conventions regarding the spherical harmonics in Appendix~\ref{sec:sphericalharm}. 

The D1, D5 charges $Q_1$, $Q_5$ have been defined in \eqref{eq:Q1Q5}; $Q_p$ represents the momentum charge and is quantized in terms of the integer $n_p$ as
\begin{equation}
Q_p = \frac{(2\pi)^4 \,n_p \,g_s^2\,\alpha'^4}{R_y^2 \,V_4}\,.
\end{equation}
By an appropriate choice of the $\mathbb{R}^4$ origin, one can choose
\begin{equation}\label{gaugecondition}
f^1_{1(\alpha,\dot \alpha)} +f^5_{1(\alpha,\dot \alpha)}=0 \quad \mathrm{for}\quad \alpha,\dot \alpha= \pm\,,
\end{equation}
which completely fixes the coordinate system (for notational convenience we use the indices $(\alpha,\dot\alpha)=(\pm,\pm)$ instead of $(m_1,\bar m_1)=(\pm 1/2,\pm 1/2)$ for $k=1$). At the first non-trivial order, one thus has the independent coefficients $f^1_{1 (\alpha,\dot\alpha)}$, $\mathcal{A}_{1(\alpha,\dot\alpha)}$ and $a_{a\pm}$, and these encode the expectation values of the dimension 1 operators $\Sigma_2^{\alpha \dot\alpha}$, $O^{\alpha \dot\alpha}$, $J^a$, and $\tilde J^a$. 

In the CFT we will mostly use null coordinates on the cylinder, which we also denote by $(u,v)$, and which are related to the CFT time and spatial coordinates analogously to the corresponding spacetime coordinate relations \eq{eq:ty}.
All the CFT one-point functions in this paper will consist of a light operator $O_i$ inserted at a generic point $(u,v)$ in the background of a heavy state:
\be
\langle O_i \rangle \;\equiv\; \langle H | O_i(u,v) | H \rangle \,.
\ee
The dependence on the insertion point $(u,v)$ is determined by conformal invariance, and in fact the expectation values of the operators we consider in RR ground states are independent of $(u,v)$ and are controlled solely by the zero mode of the light operator $O_i$. For the superstratum states that we shall study in Section \ref{sec:superstrata}, some of the one-point functions will however have non-trivial $v$ dependence.

When the expectation value is taken in the heavy state dual to the geometry corresponding to \eqref{eq:geometryexpansion},
the precise map\footnote{\label{foot:sign}The term $(-1)^{\alpha\dot{\alpha}}$ gives a minus sign when $(\alpha,\dot{\alpha})=(\pm,\mp)$. This is required by $SU(2)_L\times SU(2)_R$ invariance: the scalar product between two operators $\mathcal{O}_1$ and $\mathcal{O}_2$ with indices in the fundamental of $SU(2)_L\times SU(2)_R$ is given by $\mathcal{O}_1\cdot \mathcal{O}_2=\epsilon_{\alpha\beta}\epsilon_{\dot{\alpha}\dot{\beta}}\mathcal{O}_1^{\alpha\dot{\alpha}}\mathcal{O}_2^{\beta\dot{\beta}}$.} is \cite{Kanitscheider:2006zf,Kanitscheider:2007wq,Giusto:2015dfa}
\begin{equation}\label{eq:mapdim1}
\begin{aligned}
\frac{\sqrt{2}}{N} \langle \Sigma_2^{\alpha \dot\alpha}\rangle = (-1)^{\alpha\dot{\alpha}}\,2\, \sqrt{\frac{N}{Q_1 Q_5}} R_y\,f^1_{1 (-\alpha,-\dot\alpha)}\quad &,\quad \frac{1}{\sqrt{N}} \langle O^{\alpha \dot\alpha}\rangle = (-1)^{\alpha\dot{\alpha}}\,2 \,\sqrt{\frac{N}{Q_1 Q_5}} R_y\,\mathcal{A}_{1 (-\alpha,-\dot\alpha)}\,,\\
\frac{1}{\sqrt{N}} \langle J^\pm \rangle =\sqrt{2}\sqrt{\frac{N}{Q_1 Q_5}} R_y\, a_{\mp,+}\quad &,\quad \frac{1}{\sqrt{N}} \langle \tilde J^\pm \rangle =\sqrt{2}\sqrt{\frac{N}{Q_1 Q_5}} R_y\, a_{\mp,-}\,, \\
\frac{1}{\sqrt{N}} \langle J^3 \rangle =\sqrt{\frac{N}{Q_1 Q_5}} R_y\, a_{0,+}\quad &,\quad \frac{1}{\sqrt{N}} \langle \tilde J^3 \rangle =\sqrt{\frac{N}{Q_1 Q_5}} R_y\, a_{0,-}\,,
\end{aligned}
\end{equation}
where  the numerical factors have been chosen in such a way that the operators on the left-hand side have unit norm in the large $N$ limit.
As anticipated below Eq.\;\eq{eq:jtildepm}, our (standard) choice of normalization of $J^a$, $\tilde{J}^a$ introduces different coefficients for $J^\pm$ and $J^3$ in this dictionary. Taking into account that the correctly normalized descendant of $J^+$ is $-\sqrt{2}J^3$, and likewise for $\tilde{J}^a$, the above expressions indeed respect the SU(2)$_L$ and SU(2)$_R$ R-symmetries.

\subsection{An example}
Several non-trivial tests of the map \eqref{eq:mapdim1} have already been performed in \cite{Giusto:2015dfa}. We present here one further example, which concentrates on the expectation values of $J^a$ and $\tilde J^a$, because it will justify the choice of sign for $a_k^{(-+)}$ in \eqref{eq:FPcurve}; this sign will be relevant in testing the map for dimension two operators.

Consider the state
\begin{equation}\label{eq:psiABC}
\sum_{p,q} (A\,\ket{++}_1)^{N-p-q}\, (B\,\ket{+-}_1)^p \,(C\,\ket{-+}_1)^q\,.
\end{equation}
From Eq.\;\eqref{eq:FPcurve}, the profile function associated to this state has the following components:
\begin{equation}\label{eq:profileabc}
g_1(v')+ig_2(v')=\bar a\,e^{\frac{2\pi i}{L}v'}\, , \quad~~
g_3(v')+ig_4(v')=\bar b\,e^{\frac{2\pi i}{L}v'}-{c}\,e^{-\frac{2\pi i}{L}v'}\, , \quad~~ g_5(v')=0 \,.
\end{equation}
This profile encodes the data needed to generate the dual geometry through Eq.\;\eqref{eq:metricdata}: since we are interested in the expectation values of the left and right currents, it follows from Eq.\;\eqref{eq:mapdim1} that the coefficients we need are
\begin{equation}\label{coefficient geometry ++1+-1-+1}
\begin{split}
a_{++}&=\frac{R_y}{\sqrt{Q_1Q_5}} \frac{{a}\bar c}{\sqrt{2}} \,,\hspace{7.35em}
a_{-+}=\frac{R_y}{\sqrt{Q_1Q_5}} \frac{\bar a {c}}{\sqrt{2}} \,,\\
a_{--}&=\frac{R_y}{\sqrt{Q_1Q_5}} \frac{\bar a {b}}{\sqrt{2}} \,,\hspace{7.35em}
a_{+-}=\frac{R_y}{\sqrt{Q_1Q_5}} \frac{{a} \bar b}{\sqrt{2}} \,,\\
a_{0+}&=\frac{R_y}{\sqrt{Q_1Q_5}}\frac{|a|^2+|b|^2-|c|^2}{2} \,,\qquad
a_{0-}=\frac{R_y}{\sqrt{Q_1Q_5}}\frac{|a|^2-|b|^2+|c|^2}{2} \,.
\end{split}
\end{equation}

The zero-mode of the CFT operator $J^3$, i.e.~$J^3_0$, has eigenvalue $1/2$ on the strands $\ket{++}_1$ and $\ket{+-}_1$ while it has eigenvalue $-1/2$ on the strands of type $\ket{-+}_1$. Since each component of the superposition in \eq{eq:psiABC} is an eigenstate of $J^3_0$, its expectation value is controlled by the average number of strands of each type:
\begin{equation}\label{eq:VEVJ3++1+-1-+1} 
\langle J^3 \rangle=\frac{1}{2}\big(\bar{N}^{++}+\bar{N}^{+-}-\bar{N}^{-+}\big)=\frac{1}{2}\frac{R_y^2\,N}{Q_1Q_5}\Big(|a|^2+|b|^2-|c|^2\Big) \,,
\end{equation}
where we have used Eqs.\;\eqref{eq:average} and~\eqref{eq:Aamap}.
Analogously one can compute the expectation value of the operator $\tilde{J}^3$, which gives
\begin{equation}\label{eq:VEVtildeJ3++1+-1-+1}
\langle \tilde{J}^3 \rangle=\frac{1}{2}\big(\bar{N}^{++}-\bar{N}^{+-}+\bar{N}^{-+}\big)=\frac{1}{2}\frac{R_y^2\,N}{Q_1Q_5}\Big(|a|^2-|b|^2+|c|^2\Big) \,.
\end{equation}
Let us now consider the operator $J^+$. Its zero-mode, $J^+_0$, maps a strand of type $\ket{-+}_1$ into $\ket{++}_1$; the strand $\ket{+-}_1$ is annihilated, so is just a spectator. Thus the expectation value is determined by the following process (here and in similar expressions, to lighten the notation we suppress the subscript $0$ and it should be understood that we are considering the zero mode of the operator, since this is the only mode that contributes to the correlator for RR ground states):
\begin{equation}
J^+ \left(\ket{++}_1^{N-p-q}\ket{+-}_1^{p}\ket{-+}^q\right)=(N-p-q+1)\left(\ket{++}_1^{N-p-q+1}\ket{+-}_1^{p}\ket{-+}^{q-1}\right).
\end{equation}
Here the factor $N-p-q+1$ arises from observing that $J^+$ can transform any of the $q$ strands of type $\ket{-+}_1$ and imposing that the total number of terms on the left and right-hand sides of the equation match.
(We will explain similar steps in more detail in Section \ref{sec:o2}).
 Thus we obtain
\begin{equation}\label{eq:VEVJ+++1+-1-+1}
\langle J^+\rangle=\frac{C}{A}(N-\bar{p}-\bar{q})=\frac{R_y^2\,N}{Q_1Q_5}\bar{a}c \,.
\end{equation}
Using $J^-=(J^+)^\dagger$, we have
\begin{equation}\label{eq:VEVJ-++1+-1-+1}
\langle J^-\rangle=\langle J^+\rangle^*=\frac{R_y^2\,N}{Q_1Q_5}a\bar{c} \,.
\end{equation}
Analogously we obtain
\begin{equation}\label{eq:VEVtildeJ+++1+-1-+1}
\langle \tilde{J}^+\rangle=\frac{B}{A}(N-\bar{p}-\bar{q})=\frac{R_y^2\,N}{Q_1Q_5}\bar{a}b \,,\quad
\langle \tilde{J}^-\rangle=\langle \tilde{J}^+\rangle^*=\frac{R_y^2\,N}{Q_1Q_5}a\bar{b} \,.
\end{equation}
Comparing the gravity coefficients in Eq.\;\eqref{coefficient geometry ++1+-1-+1} and the CFT results in Eqs.\;\eqref{eq:VEVJ3++1+-1-+1}--\eqref{eq:VEVtildeJ+++1+-1-+1}, one can verify the consistency of~\eqref{eq:FPcurve}, \eq{eq:Aamap} and~\eqref{eq:mapdim1}.

\section{D1-D5 holography at dimension two}
\label{sec:dim2}

Deriving the holographic map for operators of total dimension two involves two new levels of complication. First, as pointed out in \cite{Kanitscheider:2006zf}, not all operators are distinguished by their quantum numbers, and the map between the operator expectation values and the coefficients obtained from the asymptotic expansion of the geometry \eqref{eq:geometryexpansion} may involve a non-trivial mixing matrix. The mixing matrix was subsequently derived in \cite{Taylor:2007hs}, and our explicit tests confirm this result. Second, single-trace dimension-two operators can also mix with ``double-trace" operators given by sums of products of dimension-one operators evaluated on different CFT copies. This possibility was also discussed in \cite{Taylor:2007hs}, however the precise structure of the mixing was not worked out in full detail. 

In this section we derive the full explicit holographic dictionary for all single and double-trace operators of dimension $(h,\bar h) = (1,1)$. 
We choose to study operators of dimension (1,1) as it is for these operators that the mixing is most non-trivial, and because these operators enable us to perform new precision holographic tests of superstrata. It should be straightforward to generalize our work to perform a similar analysis of the other operators of total dimension two; such operators are however beyond the scope of this paper.

Single-trace operators in a symmetric product orbifold CFT are operators that involve a single sum over copies of the CFT (the `trace' is over the discrete gauge group $S_N$).
We begin by describing the single-trace CPOs of dimension (1,1), which are as follows:
\begin{itemize}
\item An operator of twist three,
\begin{equation}\label{eq:sigma3}
\Sigma_3^{++}\;=\;\sum_{r<s<t} (\sigma^{++}_{(rst)}+ \sigma^{++}_{(rts)})\,, \qquad\quad
\sigma^{++}_{(rst)} \;\equiv\;\tilde{J}^+_{-\frac13}J^+_{-\frac13}\sigma_{(rst)}
\end{equation}
where it should be understood that the fractional moded operators in the definition of the chiral primary $\sigma^{++}_{(rst)}$ are those associated with the permutation $(rst)$; more details can be found in Appendix~\ref{sec:twistfieldcorr}.
\item An operator of twist two,
\begin{equation}\label{eq:O2op}
O^{++}_2 \,\equiv\; \sum_{r<s} O^{++}_{(rs)}\,, \qquad\quad O^{++}_{(rs)}\;\equiv\; \big(O^{++}_{(r)}+O^{++}_{(s)}\big)\,\sigma^{++}_{(rs)}\,.
\end{equation}
Here $O^{++}_{(rs)}$ is the operator (of unit norm) that joins or splits the copies $r$ and $s$ and raises the spin by $(1/2,1/2)$; for example, when acting on copies $1$ and $2$:
\begin{equation}\label{eq:defO12}
O^{++}_{(12)} \,\ket{--}_1^2 = \ket{00}_2\;,\qquad\quad O^{++}_{(12)} \,\ket{00}_2 = \ket{++}_1^2\;.
\end{equation}
As we discussed for the operator $O^{++}$ below \eq{eq:opO}, there are $h^{1,1}(\mathcal{M})$ similar operators, and we focus on the one that obtains non-zero expectation values in the states we consider.
\item An operator in the untwisted sector,
\begin{equation}\label{eq:Omega}
\Omega^{++}\;=\;\sum_r \psi^{+1}_{(r)} \psi^{+2}_{(r)} \tilde \psi^{+1}_{(r)} \tilde \psi^{+2}_{(r)} \;=\; \sum_r J^+_{(r)} \tilde J^+_{(r)}\,.
\end{equation}
\end{itemize}
As usual one can also consider the global $SU(2)_L\times SU(2)_R$ descendants of these CPOs: the multiplet of $\Sigma_3^{++}$ will be denoted by $\Sigma^{a \dot a}_3$ with $a,\dot a=+,0,-$, and analogously for the other operators. We define the descendants to have the same norm as the highest weight state, thus for example $\Omega^{0+} = \frac{1}{\sqrt{2}} \,[J_0^-,\Omega^{++}]=-\sqrt{2}\, \sum_r J^3_{(r)} \tilde J^+_{(r)}$ and $\Omega^{00} = 2 \sum_r J^3_{(r)} \tilde J^3_{(r)}$.

As mentioned above, double-trace operators also play an important role: they are defined by taking products of single-trace operators acting on disconnected subsets of the $N$ copies. The double-trace operators with dimension $(1,1)$ are
\begin{equation}\label{eq:doubletraces}
\begin{aligned}
&(\Sigma_2 \cdot \Sigma_2)^{++}\;\equiv\; \frac{2}{N^2} \sum_{(r<s)\neq (p<q)} \sigma_{(rs)}^{++} \sigma_{(pq)}^{++}\,,\qquad~~  (J \cdot \tilde J)^{++}\;\equiv\;\frac{1}{N} \sum_{r\not = s} J^{+}_{(r)} \tilde J^{+}_{(s)}\,,\\
&~ (\Sigma_2 \cdot O)^{++}\;\equiv\;\frac{\sqrt{2}}{N^{3/2}} \sum_{\substack{r<s\\t\neq r,s}} \sigma^{++}_{(rs)} O^{++}_{(t)}\,,\qquad\qquad (O \cdot O)^{++}\;\equiv\; \frac{1}{N} \sum_{r\not = s} O^{++}_{(r)} O^{++}_{(s)}~
\end{aligned}
\end{equation}
and descendants thereof; we have chosen the $N$-dependent factors to normalize the operators. The constraints in the sum defining the double-trace $(\Sigma_2 \cdot \Sigma_2)$ mean that we are summing over all couples of pairs that have no indices in common and where, in each pair, the first entry is smaller than the second one. 

On the gravity side, the asymptotic expansion of the metric \eqref{eq:geometryexpansion} gives, at the next order in $1/r$, the set of coefficients $f^1_{2 \,I}$, $f^5_{2 \,I}$ and $\mathcal{A}_{2\,I}$, where for brevity $I\equiv (a,\dot a)$ with $a,\dot a=+,0,-$. These coefficients must be related to the expectation values of the three single-trace CPOs in \eqref{eq:sigma3}, \eqref{eq:O2op}, \eqref{eq:Omega}, eventually mixed with the double-traces in \eqref{eq:doubletraces}. Since the operator $O_2$ is in fact part of a set of $h^{1,1}(\mathcal{M})$ operators, it is natural to assume that it does not mix with the other two, and that the associated gravity coefficient is $\mathcal{A}_{2\,I}$; the quantum numbers related with $\mathcal{M}$-rotations suggest that $O_2$ may mix with the double-trace $(\Sigma_2 \cdot O)$. We will examine this simple subset in the next subsection. A more intricate and interesting structure involves $\Sigma_3$, $\Omega$ and the remaining double-traces $(\Sigma_2\cdot \Sigma_2)$, $(J\cdot \tilde J)$, $(O\cdot O)$ in \eqref{eq:doubletraces}. This will be the focus of Section~\ref{sec:sigmaomega}. 

\subsection{The operator $O_2$}
\label{sec:o2}
On the gravity side, the only relevant coefficient in this sector is $\mathcal{A}_{2\,(a,\dot a)}$; on the CFT side, this should be mapped to the expectation value of $O_2^{a\dot a}$, with a possible mixing with the double-trace $(\Sigma_2\cdot O)$:
\begin{equation}\label{eq:mapdim2O2}
\frac{\sqrt{2}}{N} \langle O_2^{a\dot a}\rangle + c_1 \langle (\Sigma_2\cdot O)^{a\dot a} \rangle=(-1)^{a+\dot a}\, \gamma\,\mathcal{A}_{2\,(-a,-\dot a)}\,,
\end{equation}
where the sign $(-1)^{a+\dot a}$ is needed for $SU(2)_L\times SU(2)_R$ invariance, as one can understand following the same logic explained in Footnote~\ref{foot:sign}.
We will determine the coefficients $\gamma$ and $c_1$ by calibrating the map \eqref{eq:mapdim2O2} using some appropriately chosen RR ground states. Tests of this map will be performed in Section~\ref{sec:superstrata}, by comparing with some three-charge superstratum states.

A set of states in which  $O_2^{--}$ and $O_2^{++}$ have a non-vanishing expectation value is
\begin{equation}\label{eq:psi1O2}
\sum_{p=1}^{N/2} (A_1 \ket{++}_1)^{N-2p} (B_1 \ket{00}_2)^p\,.
\end{equation}
This expectation value can be computed following the general logic explained in \cite{Giusto:2015dfa}, which we now briefly review. Acting on two chosen strands of type $\ket{++}_1$, (the zero mode of) $O_2^{--}$ joins them into the strand $\ket{00}_2$:
\begin{equation}
O_2^{--} \,\ket{++}_1^2  \;=\; \ket{00}_2\,.
\end{equation}
When acting on the full state $(\ket{++}_1)^{N-2p}  (\ket{00}_2)^p$, there are $\binom{N-2p}{2}$ ways to choose two out of $N-2p$ strands $\ket{++}_1$; one should also take into account that the states $\psi_{\{ N_k^{(s)}\}}$ defined in \eqref{eq:RRstate} are composed of $\left |\psi_{\{ N_k^{(s)}\}}\right |^2$ terms, with $\left |\psi_{\{ N_k^{(s)}\}}\right |^2$ given in \eqref{eq:norm}. This leads to
\begin{equation}\label{eq:VEVO21part}
O_2^{--}\left(\ket{++}_1^{N-2p}  \ket{00}_2^p\right)\;=\; (p+1)\,\ket{++}_1^{N-2p-2}  \ket{00}_2^{p+1}\,,
\end{equation}
where the factor $p+1$ is the one needed to match the number of terms on the two sides of the equation, since
\begin{equation}
\binom{N-2p}{2}\, \left | (\ket{++}_1)^{N-2p} (\ket{00}_2)^p \right |^2  \;=\; (p+1)\,\left | (\ket{++}_1)^{N-2p-2} (\ket{00}_2)^{p+1} \right |^2  \,.
\end{equation}
The expectation value of $O_2^{--}$ in the state \eq{eq:psi1O2} then follows from \eqref{eq:VEVO21part} and the definition of the state \eqref{eq:psi1O2}:
\begin{equation}\label{eq:OmmCFT}
\langle O_2^{--} \rangle \;=\; \frac{A_1^2}{B_1}\, \overline{p+1} \;\approx\; \frac{A_1^2}{B_1}\, \overline{p} \;=\;  \frac{A_1^2 \,\bar B_1}{2} \,,
\end{equation}
where we have taken the large $N$ (and large $p$) limit and used \eqref{eq:average}. On the gravity side the state \eqref{eq:psi1O2} is dual to the D1-D5 geometry associated with the profile
\begin{equation}
g_1(v')+i g_2(v')\,=\,\bar a_1\,e^{\frac{2\pi i}{L}v'}\,, \quad~~ g_3(v')\,=\, g_4(v')\,=\,0\,\,,\quad~~ g_5(v')\,=\,-\mathrm{Im}\left(\frac{\bar b_1}{2}\,e^{\frac{4\pi i}{L}v'}\right),
\end{equation}
with the $a_1$, $b_1$ parameters linked to $A_1$, $B_1$ by \eqref{eq:Aamap}. Using the definition of $Z_4$ in \eqref{eq:metricdata} it is immediate to extract from the expansion \eqref{eq:geometryexpansion} the coefficients $\mathcal{A}_{2 \,(a,\dot a)}$:
\begin{equation}\label{eq:OmmGrav}
\mathcal{A}_{2\, (+,+)}\;=\;\left( \mathcal{A}_{2 \,(-,-)}\right )^*\;=\; \frac{R_y}{2\sqrt{3} \,(Q_1 Q_5)^{1/2}} \,a_1^2 \,{\bar b}_1\;=\; \frac{Q_1 Q_5}{N^{3/2} R_y^2}\,\frac{A_1^2 \,{\bar B}_1}{\sqrt{6}}\,.
\end{equation}
Note that $\mathcal{A}_{1 \,(\alpha,\dot \alpha)}=0$, consistently with the fact that the expectation value of $O^{\alpha,\dot\alpha}$ in the state \eqref{eq:psi1O2} vanishes.
Comparing the CFT \eqref{eq:OmmCFT} and gravity \eqref{eq:OmmGrav} results with the general map \eqref{eq:mapdim2O2}, one determines the parameter $\gamma$:
\begin{equation}\label{gamma map O2}
\gamma\;=\;\sqrt{3}\,\frac{N^{1/2}\,R_y^2}{Q_1 Q_5}\,.
\end{equation}

To fix the coefficient $c_1$ we must consider a state with a non-vanishing expectation value for the double-trace $(\Sigma_2 \cdot O)^{++}$. An example is
\begin{equation}\label{eq:A2B2C2OS}
\sum_{q=1}^{N/2}\sum_{p=1}^{N-2q} (A_2 \ket{++}_1)^{N-p-2q} (B_2 \ket{00}_1)^{p} (C_2 \ket{++}_2)^{q} \,.
\end{equation}
The geometry associated with this state is sourced by the following profile:
\begin{equation}
g_1(v')+i g_2(v')=\bar a_2\,e^{\frac{2\pi i}{L}v'}+\frac{\bar c_2}{2}\,e^{\frac{4\pi i}{L}v'}\,, \quad g_3(v')= g_4(v')=0\,\,,\quad g_5(v')=-\mathrm{Im}\left(\bar b_2\,e^{\frac{2\pi i}{L}v'}\right).
\end{equation}
Choosing coordinates in which~\eqref{gaugecondition} is satisfied and using~\eqref{eq:metricdata}, we obtain that the coefficient encoding the expectation value of $(\Sigma_2 \cdot O)$ takes the following value for this microstate:
\begin{equation}\label{coeff geom ++1001++2}
\mathcal{A}_{2\,(1,1)}\;=\;\frac{R_y^3}{(Q_1Q_5)^{3/2}}\,\frac{\bar{a}_2(b_2^3\,\bar{c}_2-8\bar{a}_2^2\,b_2\,c_2)}{16\sqrt{3}}\,.
\end{equation}
We now consider the action of $(\Sigma_2 \cdot O)^{++}=\sqrt{2}N^{-3/2}\Sigma_2^{++}O^{++}$ on the state~\eqref{eq:A2B2C2OS}. The operator $O^{++}$ contributes via the basic process $O^{++}\ket{00}_1=\ket{++}_1$, so that we have:
\begin{equation}\label{eq:process Sigma2O double 1}
(\Sigma_2 \cdot O)^{++}\big(\ket{++}_1^{N-p-2q}\ket{00}_1^{p}\ket{++}_2^{q}\big)=\frac{\sqrt{2}}{N^{3/2}}\Sigma_2^{++}(N-p-2q+1)\big(\ket{++}_1^{N-p-2q+1}\ket{00}_1^{p-1}\ket{++}_2^{q}\big)\,,
\end{equation}
where the factor $(N-p-2q+1)$ arises from imposing that the number of terms on the two sides of the equation match, after taking into account that the operator $O^{++}$ can act on any of the $p$ strands of type $\ket{00}_1$. The action of the operator $\Sigma_2^{++}$ is slightly more complicated: its expectation value receives a contribution both by the splitting a strand of type $\ket{++}_2$ into two $\ket{++}_1$ and from the joining of two $\ket{++}_1$ to form a $\ket{++}_2$. We thus have to consider the following basic processes (as before, the zero mode should be understood):
\begin{equation} \label{eq:use-app-1}
\Sigma_2^{++}\ket{++}_2=\ket{++}_1\ket{++}_1\,\,,\qquad
\Sigma_2^{++}\ket{00}_1\ket{00}_1=-\frac{1}{4}\ket{++}_2 \;,
\end{equation}
where the coefficient of the latter process is computed in Appendix~\ref{sec:twistfieldcorr}, see Eq.\;\eq{eq:app-b-result}. Continuing from Eq.\;\eqref{eq:process Sigma2O double 1}, we obtain
\begin{equation}
\begin{split}
(\Sigma_2 \cdot O)^{++}(&\ket{++}_1^{N-p-2q}\ket{00}_1^{p}\ket{++}_2^{q})~=~ 
\frac{\sqrt{2}}{N^{3/2}}(N-p-2q+1) \qquad \\
&~~~~~~~\left[\frac12
(N-p-2q+2)(N-p-2q+3)(\ket{++}_1^{N-p-2q+3}\ket{00}_1^{p-1}\ket{++}_2^{q-1}) \right. \\&
\left.~~~~~~~~~~~~~{}-\frac{q+1}{4}(\ket{++}_1^{N-p-2q+1}\ket{00}_1^{p-3}\ket{++}_2^{q+1})
\right]\,,
\end{split}
\end{equation}
where the combinatorial factors again arise from matching the norms of the states on both sides of the equation.
In the large $N$ limit, this gives rise to the one-point function:
\begin{equation}\label{eq:VeV Sigma2O double 1}
\big\langle (\Sigma_2 \cdot O)^{++}\big\rangle \;=\;\frac{\sqrt{2}}{N^{3/2}} \Big(\frac{\bar{A}_2^3\,B_2\,C_2}{2}-\frac{\bar{A}_2\,B_2^3\,\bar{C}_2}{8}\Big)\;=\;\frac{R_y^5N}{(Q_1Q_5)^{5/2}}\Big(\frac{\bar{a}_2^3\,b_2\,c_2}{2}-\frac{\bar{a}_2\,b_2^3\,\bar{c}_2}{16}\Big)\,,
\end{equation}
where we have used Eqs.\;\eqref{eq:windingconstraintbis} and~\eqref{eq:Aamap}. By comparing the results in Eqs.\;\eqref{coeff geom ++1001++2} and~\eqref{eq:VeV Sigma2O double 1} and the map~\eqref{eq:mapdim2O2}, we determine the unknown coefficient to be
\begin{equation}\label{c1 map O2}
c_1\,=\,-\frac{1}{N^{1/2}}\,.
\end{equation}

The holographic map in this subsector can then be summarized as
\begin{equation}\label{eq:mapdim2O2final}
\frac{\sqrt{2}}{N} \big\langle \medtilde O_2^{a\dot a}\big\rangle \;=\;(-1)^{a+\dot a}\, \sqrt{3}\,\frac{N^{1/2}\,R_y^2}{Q_1 Q_5}\,\mathcal{A}_{2\,(-a,-\dot a)}\,,
\end{equation}
where
\begin{equation}\label{eq:deftildeO2}
\medtilde O^{++}_2 \;\equiv\;  \sum_{r<s} O^{++}_{(rs)} - \frac{1}{N} \sum_{\substack{r<s\\t\neq r,s}} \sigma^{++}_{(rs)} O^{++}_{(t)} \;.
\end{equation}

By general arguments, extremal three-point functions containing the operator $\medtilde O^{++}_2$ should vanish~\cite{DHoker:1999jke,Arutyunov:2000ima,Kanitscheider:2006zf,Taylor:2007hs,Rastelli:2017udc}. We can use this as a consistency check of our result. Consider for example the correlator
\begin{equation}\label{eq:extremalO2}
\big\langle \medtilde O^{++}_2\, O^{--} \,\Sigma^{--}_2 \big\rangle = \frac{N^2}{2} \big\langle O^{++}_{(12)} \,(O^{--}_{(1)}+O^{--}_{(2)})\,\sigma^{--}_{(12)} \big\rangle -\frac{N^2}{2}\,,
\end{equation}
where the first term on the right-hand side comes from the single-trace part of $\medtilde O^{++}_2$ and the second term is produced by the double-trace part. The definition of $O^{++}_{(12)}$, \eqref{eq:defO12}, implies that 
\begin{equation}
\big\langle O^{++}_{(12)} \,(O^{--}_{(1)}+O^{--}_{(2)})\,\sigma^{--}_{(12)} \big\rangle=1\,,
\end{equation}
and thus the extremal correlator \eqref{eq:extremalO2} vanishes.

\section{The operators $\Sigma_3$ and $\Omega$}
\label{sec:sigmaomega}

In this section we turn to the sector of dimension (1,1) operators that contains $\Sigma_3$ and $\Omega$, in which the mixing is more involved. We begin this section by importing the results of \cite{Kanitscheider:2006zf,Kanitscheider:2007wq} that for a metric of the form \eqref{eq:metric6D}, with the choice of coordinates defined by \eqref{eq:polarcoord} and \eq{gaugecondition},  the geometric quantities dual to the operator expectation values in this sector are linear combinations of the following gauge-invariant quantities (evaluated in this gauge)~\cite[Eq.\;(6.4)]{Kanitscheider:2006zf},\cite[Eq.\;(5.27)]{Kanitscheider:2007wq}:
\begin{equation}\label{eq:gtildeg}
g_I \,\equiv\, \sqrt{6} \,(f^1_{2 \,I} - f^5_{2 \,I})\,,\qquad \tilde g_I \;\equiv\; \sqrt{2}\,(- (f^1_{2 \,I} + f^5_{2 \,I}) + 8 \, a_{a +} a_{b -}\,f_{I a b})\,,
\end{equation}
where the coefficients $f_{I ab}$ are defined by the overlap between a scalar S$^3$ spherical harmonic of degree 2 and the scalar product of two vector spherical harmonics of degree 1, and are given in Appendix \ref{sec:sphericalharm}.

A first guess for the holographic dictionary might have been that $g_I$ should be dual to the expectation value of $\Sigma_3^{-I}$ and $\tilde g_I$ should be dual to the expectation value $\Omega^{-I}$, however in \cite{Kanitscheider:2006zf} it was pointed out that this guess was inconsistent with the structure of known CFT correlators, and a modified map was proposed in \cite{Taylor:2007hs}. In what follows we shall not assume any previous results on the holographic dictionary beyond \eq{eq:gtildeg}, and we shall simply start with the most general map, allowing for generic mixings with the double-traces that can mix with $\Sigma_3$ and $\Omega$:
\begin{equation}\label{eq:mapdim2}
\begin{aligned}
\!\!\frac{\sqrt{3}}{N^{3/2}}\big\langle\Sigma_3^{a \dot a}\big\rangle+ a_1 \big\langle(J \cdot \tilde J)^{a \dot a}\big\rangle+a_2 \big\langle(\Sigma_2 \cdot \Sigma_2)^{a \dot a}\big\rangle+a_3 \big\langle(O \cdot O)^{a \dot a}\big\rangle&=(-1)^{a+\dot a}\Big[\alpha\,g_{(-a,-\dot a)}+\tilde \alpha\,\tilde g_{(-a,-\dot a)}\Big] ,~\\
\!\!\frac{1}{N^{1/2}}\langle\Omega^{a \dot a}\rangle+ b_1 \langle(J \cdot \tilde J)^{a \dot a}\rangle+b_2 \langle(\Sigma_2 \cdot \Sigma_2)^{a \dot a}\rangle+b_3 \langle(O \cdot O)^{a \dot a}\rangle&=(-1)^{a+\dot a}\left [\beta\,g_{(-a,-\dot a)}+\tilde \beta\,\tilde g_{(-a,-\dot a)}\right] \!.~
\end{aligned}
\end{equation}
As usual the numerical factors in front of $\Sigma_3$ and $\Omega$ have the purpose of normalizing the operators, and the sign $(-1)^{a+\dot a}$ is required by $SU(2)_L\times SU(2)_R$ invariance. 

In the following, we shall determine in turn the unknown coefficients $\alpha$, $\tilde \alpha$, $\beta$, $\tilde \beta$, $a_i$ and $b_i$ by applying the holographic map to an appropriate set of D1-D5 RR ground states. Note that we have implemented $SU(2)_L\times SU(2)_R$ invariance by requiring that coefficients be independent of the R-symmetry indices $(a,\dot a)$. (The real coefficients $\alpha, \beta$ should not be confused with the one-form $\betab$ or the spinorial indices of the R-symmetry group $SU(2)_L\times SU(2)_R$ used elsewhere.) We will then perform a set of non-trivial checks of the resulting dictionary by testing it on a wider class of states. Further tests involving D1-D5-P superstrata will be performed in Section~\ref{sec:superstrata}.

\subsection{Determining the first set of coefficients}
\label{sec:alphabeta}

To determine the values of the coefficients $\alpha$, $\tilde{\alpha}$, $\beta$, $\tilde{\beta}$, we consider states in which $\Sigma_3$ and $\Omega$ have non-zero expectation values, and in which the expectation values of the double-traces in \eq{eq:mapdim2} vanish. Two simple choices are
\begin{equation}\label{eq:psi1A1B1}
\psi^{(1)}(A_{1}, B_{1}) \;=\; \sum_{p=1}^{N/3} \big(A_{1} \ket{++}_1\big)^{N-3p} \big(B_{1} \ket{++}_3\big)^p\,,
\end{equation}
and
\begin{equation}
\psi^{(2)}(A_{2},B_{2}) \;=\;  \sum_{p=1}^{N} \big(A_{2} \ket{++}_1\big)^{N-p} \big(B_{2} \,\ket{--}_1\big)^p\,,
\end{equation}
which, according to the map in Section~\ref{sec:states}, correspond respectively to the profiles
\begin{equation}
g^{(1)}_1(v')+i g^{(1)}_2(v')=\bar a_{1}\,e^{\frac{2\pi i}{L} v'}+\frac{\bar b_{1}}{3}\,e^{\frac{6\pi i}{L} v'}\,,\qquad g^{(1)}_3(v')=g^{(1)}_4(v')=g^{(1)}_5(v')=0\,,
\end{equation}
and
\begin{equation}
g^{(2)}_1(v')+i g^{(2)}_2(v')=\bar a_{2}\,e^{\frac{2\pi i}{L} v'}+{b}_{2}\,e^{-\frac{2\pi i}{L} v'}\,,\qquad g^{(2)}_3(v')=g^{(2)}_4(v')=g^{(2)}_5(v')=0\,.
\end{equation}
The computation of the gravity parameters $g_I$ and $\tilde g_I$ follows straightforwardly from Eqs.\;\eqref{eq:metricdata}, \eqref{eq:geometryexpansion} and \eqref{eq:gtildeg}; for the state $\psi^{(1)}$ one obtains
\begin{equation}
\begin{aligned}
&g^{(1)}_{(0,0)} \,=\, -6\sqrt{2}\frac{R_y^2}{Q_1 Q_5}\,|a_1|^2\, |b_1|^2\,,\qquad\qquad~~ \tilde g^{(1)}_{(0,0)} \,=\, \frac{14\sqrt{6}}{27}\frac{R_y^2}{Q_1 Q_5}\,|a_1|^2\, |b_1|^2\,,\\
&g^{(1)}_{(1,1)} \,=\,(g^{(1)}_{(-1,-1)})^* \,=\, \sqrt{2}\frac{R_y^2}{Q_1 Q_5}\,a_1^3\, {\bar b}_1\,,\qquad \tilde g^{(1)}_{(1,1)} \,=\,(\tilde g^{(1)}_{(-1,-1)})^* \,=\, -\frac{\sqrt{2}}{\sqrt{3}}\frac{R_y^2}{Q_1 Q_5}\,a_1^3 \,{\bar b}_1\,,
\end{aligned}
\end{equation}
and for the state $\psi^{(2)}$ one obtains
\begin{equation}
\begin{aligned}
&g^{(2)}_{(0,0)} \,=\, 2\sqrt{2}\frac{R_y^2}{Q_1 Q_5}\,|a_2|^2\, |b_2|^2\, ,\qquad~~~~ \tilde g^{(2)}_{(0,0)} \,=\, 2 \sqrt{6}\frac{R_y^2}{Q_1 Q_5}\,|a_2|^2\, |b_2|^2\,,\\
&g^{(2)}_{(1,1)} \,=\,g^{(2)}_{(-1,-1)})^* \,=\, -\sqrt{2}\, a_2 \,{\bar b}_2 \,,\qquad \tilde g^{(2)}_{(1,1)} \,=\,(\tilde g^{(2)}_{(-1,-1)})^* \,=\, -\sqrt{6} \,a_2 \,{\bar b}_2\,.
\end{aligned}
\end{equation}

On the CFT side, $\Sigma_3^{--}$ and $\Omega^{--}$ have non-vanishing expectation values respectively in $\psi^{(1)}$ and $\psi^{(2)}$, while the expectation values of all the double-trace operators in \eq{eq:mapdim2} with spin $(-1,-1)$ are zero, as can be easily seen from the fact that the action of the dimension-one operators $\Sigma_2^{--}$, $J^-$, $\tilde J^{-}$ or $O^{--}$ on either $\psi^{(1)}$ or $\psi^{(2)}$ would produce strands of a type that is not present in the state itself. 

The expectation value of $\Sigma_3^{--}$ in $\psi^{(1)}$ arises from the process in which three strands of winding one are joined into a strand of winding three. In general one has (as before the zero mode should be understood here and in similar equations that follow)
\be
\sigma_{(3)}^{--}
 \ket{++}_{k_1}  \ket{++}_{k_2}  \ket{++}_{k_3} \,=\, c_{k_1, k_2, k_3}\, \ket{++}_{k_1+k_2+k_3}\,, 
\ee
where $(3)$ denotes a permutation that joins together the three strands $ \ket{++}_{k_i}$ and where $c_{k_1, k_2, k_3}=\frac{k_1+k_2+k_3}{3 k_1 k_2 k_3}$~\cite{Tormo:2018fnt}. 
We first focus on three particular strands of winding one and one particular permutation, say $(123)$, of the three strands, for which we thus have
\begin{equation}
\sigma_{(123)}^{--} \big(\ket{++}_1\big)^3 = \ket{++}_3\,.
\end{equation}
When considering the action of the full operator $\Sigma_3^{--}$ on the state $\ket{++}_1^{N-3p} \ket{++}_3^p$, one must also include the appropriate combinatorial factors, as follows. The twist operator can act on any three of the $N-3p$ strands of winding one, and for each choice of the three strands there are two inequivalent 3-cycles (c.f.~Eq.\;\eqref{eq:sigma3}). Thus $\Sigma_3^{--}$ can act in $2 \binom{N-3p}{3}$ ways on $(\ket{++}_1)^{N-3p} (\ket{++}_3)^p$ to produce the state $(\ket{++}_1)^{N-3p-3} (\ket{++}_3)^{p+1}$. Moreover one has to take into account that the initial and final states have a non-trivial norm given by \eqref{eq:norm}. Matching the norm of the states on both sides of the following equation, one finds
\begin{equation}
\Sigma_3^{--} \left((\ket{++}_1)^{N-3p} (\ket{++}_3)^p \right) =(p+1) \,(\ket{++}_1)^{N-3p-3} (\ket{++}_3)^{p+1}\,.
\end{equation}
The above result and the definition of the state $\psi^{(1)}$ in \eqref{eq:psi1A1B1} imply that, in the large $N$ limit. the expectation value of $\Sigma_3^{--}$ in the state $\psi^{(1)}$ is:
\begin{equation}
\langle \Sigma_3^{--} \rangle_1 \;=\; \frac{A_1^3}{B_1}\, {\bar p}\;=\;\frac{A_1^3\, {\bar B_1}}{3} \;=\; \frac{N^2\,R_y^2}{3\, (Q_1 Q_5)^2} \,a_1^3\, {\bar b_1}\,,
\end{equation}
where we have used ${\bar p} = |B|^2/3$ (from \eqref{eq:average}) and the relation \eqref{eq:Aamap}.

Next, the expectation value of $\Omega^{--}$ in the state $\psi^{(2)}$ arises from the basic process where $\Omega^{--}$ maps $\ket{++}_1$ to $\ket{--}_1$. There are $N-p$ choices of strand for $\Omega^{--}$ to act on the state $(\ket{++}_1)^{N-p} (\ket{--}_1)^p$ to give $(\ket{++}_1)^{N-p-1} (\ket{--}_1)^{p+1}$. Matching the norms of left and right-hand sides gives
\begin{equation}
\Omega^{--}  \left((\ket{++}_1)^{N-p} (\ket{--}_1)^p \right) = (p+1) \,(\ket{++}_1)^{N-p-1} (\ket{--}_1)^{p+1}\,,
\end{equation}
and thus the expectation value of $\Omega^{--}$ on $\psi^{(2)}$ is
\begin{equation}
\langle \Omega^{--} \rangle_2 \;=\; \frac{A_2}{B_2}\, {\bar p}\;=\; A_2\, {\bar B_2} \;=\; \frac{N\,R_y^2}{Q_1 Q_5} \,a_2 \,{\bar b_2}\,,
\end{equation}
where we have again used \eqref{eq:average} and \eqref{eq:Aamap}.

Comparing $\langle \Sigma_3^{--} \rangle_1$ and $\langle \Omega^{--} \rangle_2 $ with the gravity data $g^{(i)}_{-1,-1}$, $\tilde g^{(i)}_{-1,-1}$ ($i=1,2$) uniquely fixes $\alpha$, $\tilde\alpha$, $\beta$, $\tilde\beta$ to be
\begin{equation}\label{eq:aatildebbtilde}
\alpha\;=\;-\tilde \beta \;=\; \frac{\sqrt{3}}{4\sqrt{2}}\,\frac{N^{1/2}\,R_y^2}{Q_1 Q_5}\,,\qquad \tilde\alpha\;=\;\beta\;=\;-\frac{1}{4\sqrt{2}}\,\frac{N^{1/2}\,R_y^2}{Q_1 Q_5}\;.
\end{equation}
These values are in agreement with the results of~\cite{Taylor:2007hs}. The expectation values of $\Sigma_3^{++}$ and $ \Omega^{++}$ are simply the complex conjugates of the ones considered above, and do not add new information. The expectation values of $\Sigma_3^{00}$ and $ \Omega^{00}$ are also non-vanishing, and should be compared with $g^{(i)}_{0,0}$. For this value of the spin, however, double-trace operators play a role and so we will return to this comparison in Section~\ref{sec:tests}, where we will perform some non-trivial consistency checks of the full dictionary.

\subsection{Determining the coefficients $a_1$, $b_1$}
\label{sec:a1b1}
The coefficients $a_1$, $b_1$ in the general map \eq{eq:mapdim2} correspond to the double-trace operator $(J \cdot \tilde J)$.
An RR ground state in which $(J \cdot \tilde J)^{++}$ is the only operator with $j=\bar j=1$ to have non-vanishing expectation value is the state given in Eq.\;\eqref{eq:psiABC}. 
It is straightforward to compute this one-point function in the orbifold CFT, where $\tilde J^+$ can map any of the $p$ strands of type $\ket{+-}_1$ into $\ket{++}_1$, and likewise $J^+$ can act on any of the $\,q\,$ $\ket{-+}_1$ strands. Taking into account the normalization \eqref{eq:norm} of the states, one finds
\begin{equation}
\begin{aligned}
&(J \cdot \tilde J)^{++}\left(\ket{++}_1^{N-p-q}\ket{+-}_1^p \ket{-+}_1^q\right)\\
&\qquad=\frac{(N-p-q+1)(N-p-q+2)}{N} \ket{++}_1^{N-p-q+2}\ket{+-}_1^{p-1} \ket{-+}_1^{q-1}\,,
\end{aligned}
\end{equation}
and, in the large $N$ limit,
\begin{equation}
\langle (J \cdot \tilde J)^{++} \rangle = \frac{B\,C}{A^2} \,\frac{(N-{\bar p}-{\bar q})^2}{N}=\frac{{\bar A}^2\,B\,C}{N}= \frac{N\,R_y^4}{(Q_1 Q_5)^2}\,{\bar a}^2\,b\,c\,.
\end{equation}
Notice that, up to the normalization factor $N^{-1}$, the expectation value of $(J \cdot \tilde J)^{++}$ is just the product of the expectation values of $J^+$ and $\tilde J^+$, at large $N$.

On the gravity side, the relevant coefficients extracted from the metric associated with the profile \eqref{eq:profileabc} are
\begin{equation}
g_{1,1}=(g_{-1,-1})^* = \sqrt{2} \,\frac{R_y^2}{Q_1 Q_5}\,a^2 \,{\bar b}\,{\bar c}\;,\qquad \tilde g_{1,1}=(\tilde g_{-1,-1})^* = \sqrt{6} \,\frac{R_y^2}{Q_1 Q_5}\,a^2 \,{\bar b}\,{\bar c}\,,
\end{equation}
which, taking into account the values of $\alpha$, $\tilde\alpha$, $\beta$, $\tilde \beta$ derived in \eqref{eq:aatildebbtilde}, implies that
\begin{equation}
\alpha\,g_{-1,-1}+ \tilde\alpha\,\tilde g_{-1,-1}\,=\,0~,\qquad \beta\,g_{-1,-1}+ \tilde\beta\,\tilde g_{-1,-1}\,=\,-\frac{N^{1/2}\,R_y^4}{(Q_1 Q_5)^2}\,.
\end{equation}
Then comparison with \eqref{eq:mapdim2} yields
\begin{equation}
a_1=0\quad,\quad b_1 = -\frac{1}{N^{1/2}}\,.
\end{equation}

Using the above value of $b_1$, one sees that the combination appearing in the holographic map is 
\begin{equation}
\frac{1}{N^{1/2}}\,\left(\Omega^{++}  - \frac{1}{N} \sum_{r\neq s} J^+ \tilde J^+\right)\equiv \frac{1}{N^{1/2}}\,\medtilde \Omega^{++}\,.
\end{equation}
We note that the operator $\medtilde \Omega^{++}$ has the property that its extremal three-point function with $J^-$ and $\tilde J^-$ vanishes,
\begin{equation}
\langle \medtilde \Omega^{++} \,J^-\,\tilde J^- \rangle=0\,.
\end{equation}

\subsection{Determining the coefficients $a_2$, $b_2$}
\label{sec:a2b2}

The coefficients $a_2$, $b_2$ in the map \eq{eq:mapdim2} correspond to the operator $(\Sigma_2\cdot \Sigma_2)$. 
An RR ground state in which $(\Sigma_2\cdot \Sigma_2)^{--}$ is the only operator with $j=\bar j=-1$ to have non-vanishing expectation value is
\begin{equation}\label{eq:psiAB12}
\sum_{p=1}^{N/2} (A\,\ket{++}_1)^{N-2p} (B\,\ket{++}_2)^p \,.
\end{equation}
The CFT expectation value follows from the relation
\begin{equation}\label{eq:sigma2doubleCFT}
(\Sigma_2\cdot \Sigma_2)^{--} \,\left( \ket{++}_1^{N-2p} \,\ket{++}_2^p \right) = \frac{2 (p+1)(p+2)}{N^2}\,\ket{++}_1^{N-2p-4} \,\ket{++}_2^{p+2}\,;
\end{equation}
the combinatorial factor is derived by noting that the first $\sigma_2^{--}$ in the double-trace can act in $\binom{N-2p}{2}$ ways on the $N-2p$ strands $\ket{++}_1$ and similarly the second $\sigma_2^{--}$ can act in $\binom{N-2p-2}{2}$ ways on the remaining $N-2p-2$ strands $\ket{++}_1$; one then, as usual, equates the numbers of terms composing the states on the two sides of \eqref{eq:sigma2doubleCFT} and multiplies by the normalization factor $2/N^2$. The expectation value in the coherent state \eqref{eq:psiAB12}, for which $2{\bar p}=|B|^2$, is then 
\begin{equation}\label{eq:VEVS2S2}
\langle (\Sigma_2\cdot \Sigma_2)^{--} \rangle = \frac{A^4}{B^2},\frac{2\,{\bar p}^2}{N^2} = \frac{A^4\,{\bar B}^2}{2\,N^2}=\frac{N\,R_y^6}{(Q_1 Q_5)^3}\,\frac{a^4\,{\bar b}^2}{2}\,.
\end{equation}
We note that, in the large $N$ limit, the expectation value of the double-trace $(\Sigma_2\cdot \Sigma_2)^{--}$ is given again by the square of the normalized single trace $(\sqrt{2}/\sqrt{N})\:\! \Sigma_2^{--}$, which was computed in Eq.\;(4.14) of \cite{Giusto:2015dfa}.

The geometry dual to the state \eqref{eq:psiAB12} is generated from the profile
\begin{equation}
g_1(v')+i g_2(v') = \bar a\,e^{\frac{2\pi i}{L}v'}+\frac{\bar b}{2}\, e^{\frac{4\pi i}{L}v'}-\frac{R_y^2}{2\:\! Q_1 Q_5}\,\bar a^2\,b\,,\qquad g_3(v')=g_4(v')=g_5(v')=0\,,
\end{equation}
where we have shifted the profile centre in order to implement the gauge condition $f^1_{1}+f^5_{1}=0$. From this geometry one derives
\begin{equation}
g_{1,1}=(g_{-1,-1})^* = -\sqrt{2} \frac{R_y^4}{(Q_1 Q_5)^2}\,a^4\,{\bar b}^2\quad,\quad \tilde g_{1,1}=(\tilde g_{-1,-1})^* = \frac{1}{\sqrt{6}} \frac{R_y^4}{(Q_1 Q_5)^2}\,a^4\,{\bar b}^2\,.
\end{equation}
Comparing with \eqref{eq:mapdim2} and using the values \eqref{eq:aatildebbtilde}, one deduces
\begin{equation}
a_2 \;=\; -\frac{7}{4\sqrt{3}}\frac{1}{N^{1/2}}\;,\qquad b_2 \;=\; \frac{1}{4}\frac{1}{N^{1/2}}\,.
\end{equation}

\subsection{Determining the coefficients $a_3$, $b_3$}
\label{sec:a3b3}

The coefficients $a_3$, $b_3$ in the map \eq{eq:mapdim2} correspond to the double-trace operator $(O \cdot O)$.
A set of RR ground states in which $(O\cdot O)^{--}$ is the only operator with $j=\bar j=-1$ to have non-vanishing one-point function is
\begin{equation}
\sum_{p=1}^N (A\,\ket{++}_1)^{N-p} (B\,\ket{00}_1)^p\,,
\end{equation}
which is just a particular case of the state \eqref{eq:A2B2C2OS} with $C_2=0$ and $A_2= A$, $B_2= B$. The expectation value $\langle (O\cdot O)^{--} \rangle$ is, as usual, proportional to the square of the single-trace expectation value  $\langle O^{--} \rangle = A\,{\bar B}$, as computed in \cite{Giusto:2014aba}. We obtain
\begin{equation}
\langle (O\cdot O)^{--} \rangle \;=\; \frac{A^2 {\bar B}^2}{N} \;=\; \frac{N\,R_y^4}{(Q_1 Q_5)^2}\frac{a^2 \,{\bar b}^2}{2}\,.
\end{equation}
The relevant gravity coefficients are
\begin{equation}
g_{1,1}\,=\,(g_{-1,-1})^* \,=\, \frac{\sqrt{2}}{4} \frac{R_y^2}{Q_1 Q_5} \,a^2 \,{\bar b}^2\;,\qquad \tilde g_{1,1}\,=\,(\tilde g_{-1,-1})^*\,=\,- \frac{\sqrt{2}}{4\sqrt{3}} \frac{R_y^2}{Q_1 Q_5} \,a^2 \,{\bar b}^2\,,
\end{equation}
which determines $a_3$ and $b_3$ to be
\begin{equation}
a_3 \,=\, \frac{1}{2\sqrt{3}}\frac{1}{N^{1/2}}~,\qquad~~ b_3\,=\,0\,.
\end{equation}

\subsection{The holographic dictionary at dimension (1,1)}
We can now summarize our results and write the explicit holographic map in the $\Sigma_3$, $\Omega$ sector as:
\begin{equation}\label{eq:mapdim2bis}
\begin{aligned}
\frac{\sqrt{3}}{N^{3/2}}\big\langle\Sigma_3^{a \dot a}\big\rangle+\frac{1}{4\sqrt{3}}\,\frac{1}{N^{1/2}}\Big[\!{}-7\big\langle(\Sigma_2 \cdot \Sigma_2)^{a \dot a}\big\rangle+2\,\big\langle(O \cdot O)^{a \dot a}\big\rangle\Big]&\;=\; (-1)^{a+\dot a}\,h_{(-a,-\dot a)}\,,\\
\frac{1}{N^{1/2}}\big\langle\Omega^{a \dot a}\big\rangle- \frac{1}{N^{1/2}}\,\left[\big\langle(J \cdot \tilde J)^{a \dot a}\big\rangle -\frac{1}{4}\,\big\langle(\Sigma_2 \cdot \Sigma_2)^{a \dot a}\big\rangle\right]&\;=\; (-1)^{a+\dot a}\, \tilde h_{(-a,-\dot a)}\,,
\end{aligned}
\end{equation}
where (recall that $g$, $\tilde g$ were defined in \eq{eq:gtildeg})
\bea\label{eq:mapdim2data}
h_{(a,\dot a)}\,&\equiv&\, \frac{N^{1/2}\,R_y^2}{4\sqrt{2}\,Q_1 Q_5} \left[\sqrt{3}\,g_{(a,\dot a)}-\tilde g_{(a,\dot a)}\right],\cr
\tilde h_{(a,\dot a)}\,&\equiv&\, -\frac{N^{1/2}\,R_y^2}{4\sqrt{2}\,Q_1 Q_5} \left[g_{(a,\dot a)}+\sqrt{3}\,\tilde g_{(a,\dot a)}\right].
\eea
We also repeat for the reader's convenience the results from the $O_2$ sector, \eq{eq:mapdim2O2final} and \eq{eq:deftildeO2}:
\begin{equation}\label{eq:mapdim2O2final-1}
\frac{\sqrt{2}}{N} \big\langle \medtilde O_2^{a\dot a}\big\rangle \;=\;(-1)^{a+\dot a}\, \sqrt{3}\,\frac{N^{1/2}\,R_y^2}{Q_1 Q_5}\,\mathcal{A}_{2\,(-a,-\dot a)}\,,
\end{equation}
where
\begin{equation}\label{eq:deftildeO2-1}
\medtilde O^{++}_2 \;\equiv\;  \sum_{r<s} O^{++}_{(rs)} - \frac{1}{N} \sum_{\substack{r<s\\t\neq r,s}} \sigma^{++}_{(rs)} O^{++}_{(t)} \;.
\end{equation}
For the class of $\mathcal{M}$-invariant supergravity solutions with a flat four-dimensional base space, Eqs.~\eq{eq:mapdim2bis}--\eq{eq:deftildeO2-1} comprise the holographic dictionary at dimension $(1,1)$.

\newpage
One can check that not all extremal three-point functions of the operator combinations dual to $g$, $\tilde{g}$ vanish.
Based on general expectations, there should be an appropriate field redefinition such that all extremal three-point functions vanish~\cite{DHoker:1999jke,Arutyunov:2000ima,Kanitscheider:2006zf,Taylor:2007hs,Rastelli:2017udc}. We leave the determination of this field redefinition for future work.

\subsection{Tests of the holographic dictionary on two-charge states}
\label{sec:tests}
Having determined all the coefficients in the holographic map \eqref{eq:mapdim2bis}, we can now use the map as a non-trivial consistency check on the correspondence \eqref{eq:Aamap} between the 1/4-BPS RR ground states \eqref{eq:coherent} and the supergravity solutions \eqref{eq:metricdata}. We re-emphasize that the $SU(2)_L\times SU(2)_R$ symmetry requires the coefficients in \eqref{eq:mapdim2bis} to be independent of the spin $(a,\dot a)$; thus, even if the most efficient way to fix the coefficients is to focus on the highest (or the lowest) spin component, as we have done in the previous subsections, the same coefficients must necessarily reproduce the expectation values of all other components. A relatively involved example is given by the operators 
\begin{equation}
\Omega^{00}=2 \sum_r J^3_{(r)} \tilde J^3_{(r)} \qquad \mathrm{and}\qquad \Sigma_3^{00}=\frac{1}{2}\,[J_0^-,[\tilde J_0^-,\Sigma_3^{++}]]\,.
\end{equation}
We will next work out a couple of examples that demonstrate how the one-point functions of these operators are correctly reproduced by the map \eqref{eq:mapdim2bis}. More examples involving 1/8-BPS D1-D5-P states will be examined in the next section.

\vspace{2mm}
\begin{itemize}
\item First, consider the state $(A\,\ket{++}_k)^{\frac{N}{k}}$ with $k\in \mathbb{N}$.
\end{itemize}

The dual geometry is generated from the profile
\begin{equation}
g_1(v')+i g_2(v')\,=\,\frac{\bar a}{k}\,e^{\frac{2\pi i\,k}{L}v'}~,\qquad g_3(v')\,=\,g_4(v')\,=\,g_5(v')\,=\,0\,,
\end{equation}
and from the asymptotic expansion of the geometry one deduces that 
\begin{equation}
h_{(a,\dot a)}\,=\, \tilde h_{(a,\dot a)}\,=\,0\qquad~~ \mathrm{for~all}\quad (a,\dot a)\,.
\end{equation}
This is a reflection of the fact that the geometry is a $\mathbb{Z}_k$ quotient of AdS$_3\times S^3$, with non-trivial constant gauge fields mixing S$^3$ and AdS$_3$. 

Given the simple structure of the geometry, one would naively expect that on the CFT side only the R-symmetry currents, which couple to the S$^3$ gauge fields, have non-trivial expectation values; the situation is however a bit more interesting. While it is true that to leading order at large $N$ all expectation values appearing in the first line of \eqref{eq:mapdim2bis} vanish\footnote{Naively one could think that the expectation value of the double-trace $(\Sigma_2\cdot \Sigma_2)^{00}\sim \sum \sigma^{++}_{(rs)}\sigma^{--}_{(pq)}+\sigma^{+-}_{(rs)}\sigma^{-+}_{(pq)}$ could receive a contribution, for example, from the process in which a $\sigma^{--}$ joins two strands $\ket{++}_k$ into $\ket{++}_{2k}$ and a $\sigma^{++}$ splits the newly created $\ket{++}_{2k}$ strand again into two $\ket{++}_k$ strands. One can however see that this expectation value, unlike the one computed in \eqref{eq:VEVS2S2}, does not grow with $N$, and hence it does not contribute to the holographic map at the leading order in the large $N$ expansion. The origin of the difference with \eqref{eq:VEVS2S2} is that in the present situation the second twist operator can only act on a particular strand, while in \eqref{eq:VEVS2S2} it could act on $O(N)$ strands. This observation confirms the general rule that the expectation value of a double-trace operator is given by the product of the expectation values of the single-trace components at leading order in $N$.}, the expectation values of the single-trace $\Omega^{00}$ and of the double-trace $(J\cdot \tilde J)^{00}$ are non-trivial; consistency with the map \eqref{eq:mapdim2bis} requires that the two expectation values precisely cancel. To compute the expectation value of $\Omega^{00}$ one notes that 
\begin{equation}\label{eq:Omega00k}
\Omega^{00}\,\ket{++}_k = \frac{1}{2\:\! k}\,\ket{++}_k\,.
\end{equation}
The $1/k$ factor in this equation is not a-priori obvious and can be understood as follows. Consider the action of the zero-mode of the $SU(2)_L$ current $J^3_0$ on a strand of winding $k$, such as $\ket{++}_k$. Since there are identical copies of the $SU(2)_L$ algebra in any twist sector of the orbifold theory, the value of $J^3_0$ cannot depend on $k$: $J^3_0 \,\ket{++}_k = 1/2 \,\ket{++}_k$; on the other hand $J^3_0 = \sum_{r=1}^k J^3_{(r),0} = k\,J^3_{(r),0}$, with $J^3_{(r),0}$ the zero-mode of the operator acting on a single copy of the CFT. One deduces that, in the $k$-twisted sector, $J^3_{(r),0}=1/k\,J^3_0$ and analogously $\tilde J^3_{(r),0}=1/k\,\tilde J^3_0$. This implies that $\Omega^{00}_0= 2\sum_{r=1}^k J^3_{(r),0} \tilde J^3_{(r),0} = 2/k\,J^3_0 \tilde J^3_0$, from which \eqref{eq:Omega00k} immediately follows.

The action of $\Omega^{00}$ on the full state $(\ket{++}_k)^{\frac{N}{k}}$ is then given by multiplying by the number of strands $N/k$:
\begin{equation}
\Omega^{00}\,\big(\ket{++}_k\big)^{\frac{N}{k}} \,=\, \frac{N}{2\:\!k^2}\,\big(\ket{++}_k\big)^{\frac{N}{k}}\,.
\end{equation}
This immediately implies
\begin{equation}\label{eq:VEVOm00}
\big\langle \Omega^{00} \big\rangle \,=\, \frac{N}{2\:\!k^2}\,.
\end{equation}
As for the expectation value of the double-trace $(J\cdot \tilde J)^{00}$, one should first note that the correctly normalized affine descendant of $(J\cdot \tilde J)^{++}$, which is what appears in the map \eqref{eq:mapdim2bis}, is given by
\begin{equation}
(J\cdot \tilde J)^{00} \,=\, \frac{2}{N}\,\sum_{r\not = s} J^3_{(r)} \tilde J^3_{(s)} \,.
\end{equation}
When acting on the state $(\ket{++}_k)^{\frac{N}{k}}$, $J^3$ can be applied on any of the $N/k$ strands, and it has eigenvalue $1/2$. The same happens for $\tilde J^3$ on the remaining $N/k-1$ strands. In the large $N$ limit one finds
\begin{equation}
(J\cdot \tilde J)^{00} \,\big(\ket{++}_k\big)^{\frac{N}{k}} \;=\; \frac{2}{N} \frac{N^2}{k^2}\,\frac{1}{4}\,\big(\ket{++}_k\big)^{\frac{N}{k}} \;=\; \frac{N}{2\,k^2}\,(\ket{++}_k)^{\frac{N}{k}} \,,
\end{equation}
and thus
\begin{equation}\label{eq:VEVJJ00}
\big\langle (J\cdot \tilde J)^{00}  \big\rangle \,=\, \frac{N}{2\,k^2}\,.
\end{equation}
The two expectation values \eqref{eq:VEVOm00} and \eqref{eq:VEVJJ00} are equal, as required by the holographic map.

\vspace{2mm}
\begin{itemize}
\item Second, let us consider the state 
\end{itemize}
\be \label{eq:state-1}
\sum_{p=1}^{N/k} (A\,\ket{++}_1)^{N-k\,p} (B\ket{++}_k)^p \,,\qquad\quad k\in \mathbb{N}\,, \quad k\ge 3 \,.
\ee

The supergravity analysis is done along the usual lines: starting from the dual profile
\begin{equation}
g_1(v')+i g_2(v')=\bar a\,e^{\frac{2\pi i}{L}v'}+\frac{\bar b}{k}\,e^{\frac{2\pi i\,k}{L}v'}\quad,\quad g_3(v')=g_4(v')=g_5(v')=0\,,
\end{equation}
(where for simplicity we take $a,b\in \mathbb{R}$) one extracts the supergravity data defined in \eqref{eq:mapdim2data}:
\begin{equation}\label{eq:h00tildeh00}
h_{(0,0)}=\frac{\sqrt{3}}{6}\,\frac{(k+1)^2}{k^2}\,\frac{N^{1/2}\,R_y^4}{(Q_1 Q_5)^2}\,a^2 \,b^2\quad,\quad \tilde h_{(0,0)}=\frac{1}{2}\,\frac{(k-1)^2}{k^2}\,\frac{N^{1/2}\,R_y^4}{(Q_1 Q_5)^2}\,a^2 \,b^2\,.
\end{equation}
Note that in the following manipulations the regularity constraint $a^2+b^2=\frac{Q_1 Q_5}{R_y^2}$ \eqref{eq:regularityconstraint} will be used.
 
The second line of \eqref{eq:mapdim2bis} works in a way that is qualitatively similar to the previous example. We take $k\ge 3$ for simplicity, where the non-vanishing expectation values are $\langle \Omega^{00} \rangle$ and $\langle (J\cdot \tilde J)^{00}  \rangle$ (for $k=2$, one would also need to include $\langle (\Sigma_2\cdot \Sigma_2)^{00}  \rangle$). The one-point functions can be computed by applying the rules already explained:
\begin{equation}
\langle \Omega^{00} \rangle \,=\, \frac{1}{2} \,\frac{N\,R_y^4}{(Q_1 Q_5)^2}\,\frac{k^2 a^4+(k^2+1) a^2 b^2+b^4}{k^2}\;,~~~\langle (J\cdot \tilde J)^{00}  \rangle\,=\,\frac{1}{2} \,\frac{N\,R_y^4}{(Q_1 Q_5)^2}\,\frac{k^2 a^4+2k a^2 b^2+b^4}{k^2}\,.
\end{equation}
One can verify that substituting these expectation values in the second line of \eqref{eq:mapdim2bis} reproduces the value of $\tilde h_{(0,0)}$ given in \eqref{eq:h00tildeh00}.

The first line of \eqref{eq:mapdim2bis} introduces a novel ingredient: the expectation value of $\Sigma_3^{00}$ (the other double-trace operators clearly do not play a role in this example, at large $N$.). The mechanism by which $\Sigma_3^{00}$ acquires a non-zero expectation value in the state \eq{eq:state-1} for any $k>1$ is as follows. Take for example $k=3$ and consider the action of $\Sigma_3^{00}$ on the strands $\ket{++}_1$ and $\ket{++}_3$ corresponding to the permutation $(1)\,(234)$; when the twist 3 operator acts with the permutation $(132)$ it produces a state described by the permutation $(2)\,(341)$, which represents again two strands of type $\ket{++}_1$ and $\ket{++}_3$. In other words, the operator $\Sigma_3^{00}$ maps the state \eq{eq:state-1} into itself, permuting the copy $\ket{++}_1$ with one of the copies forming the strand $\ket{++}_3$. To compute the expectation value associated with this process we need to know the coefficient $C_{k,3,k}^{-,-(1),-}$ defined by
\begin{equation}\label{eq:sigma3process}
\sigma^{00}_{(3)} \,\ket{++}_1 \ket{++}_k= C_{k,3,k}^{-,-(1),-}\,\ket{++}_1 \ket{++}_k\,,
\end{equation}
where $(3)$ denotes any permutation that maps the state on the left to the state on the right. 
This coefficient is equal to $C_{3,k,k}^{-(1),-,-}$, corresponding to a three-point function that differs from the one giving $C_{k,3,k}^{-,-(1),-}$ by the ordering of the operators. One can see that the coefficients are equal using e.g.~\cite[Eq.\;(2.2.48)]{Ribault:2014hia}. The coefficient $C_{3,k,k}^{-(1),-,-}$ was computed in \cite[Eq.\;(6.28)]{Lunin:2001pw} using the techniques reviewed in Appendix~\ref{sec:twistfieldcorr}, giving
\begin{equation}
C_{k,3,k}^{-,-(1),-} = \frac{(k+1)^2}{6\,k^2}\,.
\end{equation}
The full expectation value of $\Sigma_3^{00}$ is given by dressing $C_{k,3,k}^{-,-(1),-}$ by the appropriate combinatorial factors: the twist operator can act on any of the $(N-k\,p)\,p$ pairs of strands $\ket{++}_1 \ket{++}_k$ and can cut the $\ket{++}_k$ strand in $k$ different positions (note that only one of the two permutations $(rst)$ and $(rts)$ that appear in the definition of $\Sigma_3$ \eqref{eq:sigma3} contributes to the present process, and thus one does not have an additional factor of 2). We thus find
\begin{equation}
\Sigma_3^{00}\,\ket{++}_1^{N-k\,p} \,\ket{++}_k^p \;=\; C_{k,3,k}^{-,-(1),-}\,(N-k\,p)\,p\,k \, \ket{++}_1^{N-k\,p} \,\ket{++}_k^p\,,
\end{equation}
which gives
\begin{equation}
\big\langle \Sigma_3^{00}\big\rangle \;=\; \frac{(k+1)^2}{6\,k^2}\,A^2\,B^2 \;=\;  \frac{(k+1)^2}{6\,k^2}\,\frac{N^2\,R_y^4}{(Q_1 Q_5)^2}\,a^2\,b^2\,.
\end{equation}
The CFT prediction agrees, via the map \eqref{eq:mapdim2bis}, with the gravity coefficient $h_{(0,0)}$ in \eqref{eq:h00tildeh00}.

\section{Precision holographic tests of superstrata}
\label{sec:superstrata}

We now perform new precision tests of the proposed holographic dictionary for a recently constructed set of superstratum solutions and proposed dual CFT microstates. The term `superstratum' refers to a large class of supergravity solutions describing black hole microstates~\cite{Bena:2015bea,Bena:2016agb,Bena:2016ypk,Bena:2017geu,Bena:2017upb,Bena:2017xbt,Bena:2018bbd,Bakhshaei:2018vux,Bena:2018mpb,Ceplak:2018pws,Heidmann:2019zws}. The key property of superstrata is that the isometries preserved by the black hole are explicitly broken (apart from the single null isometry guaranteed by supersymmetry). These solutions include sub-classes whose proposed dual CFT states display momentum fractionation \cite{Bena:2016agb}, and include solutions that have parametrically long AdS$_2$ throats (in full, the throats are approximately AdS$_2 \times $S$^1 \times $S$^3 \times$T$^4$) \cite{Bena:2016ypk,Bena:2017xbt}, which have potentially important implications for AdS$_2$ holography~\cite{Bena:2018bbd}. Some special sub-families have the remarkable property of having completely integrable null geodesics~\cite{Bena:2017upb}; for some recent studies of superstrata, see~\cite{Tyukov:2017uig,Bianchi:2018kzy,Bena:2018mpb,Bombini:2019vnc}.

We will perform tests on a couple of specific sub-families of superstrata, including some of the most recently constructed solutions~\cite{Ceplak:2018pws}. In all cases the proposed CFT description passes these new precision tests, which lends strong support to the proposed families of holographically dual CFT states.

\subsection{Key properties of superstrata}

We now briefly summarize the elements of the superstratum construction that will be relevant for our studies. In this paper the CFT expectation values we focus on are related to the fall-off of metric components, so for simplicity of presentation we shall focus on metric quantities. The main purpose will be to introduce the necessary notation for the holographic tests that follow. For a more comprehensive introduction to superstrata, we refer the reader to \cite{Bena:2017xbt}.

The superstrata that have been constructed to date are six-dimensional solutions where the four-dimensional base is flat $\mathbb{R}^4$. The six-dimensional metric, four-dimensional base and relation between $t,y$ and $u,v$ coordinates are as given in Eqs.\;\eq{eq:metric6D}--\eq{eq:ty}. The one-form $\betab$ takes the value
\be
\betab \;=\;  \frac{R_y a^2}{\sqrt{2}\,\Sigma}\,\big(\sin^2\theta\, d\phi - \cos^2\theta\,d\psi\big) \,. 
\ee 
The remaining quantities in the supergravity ansatz \eq{ansatzSummary} are organized by the almost-linear structure of the six-dimensional BPS equations. For completeness we give the full Type IIB ansatz and BPS equations in Appendix \ref{app:sugra}, and we summarize the content here. The four-dimensional base and the one-form $\betab$ are referred to as the data of the ``zeroth'' layer of equations. Then the first layer of BPS equations involves the scalars $Z_1$, $Z_2$, $Z_4$ and two-forms $\Theta^1$, $\Theta^2$, $\Theta^4$. By convention $Z_3$ is related to $\cF$, and $\Theta^3 = d\betab$.  Finally, the second layer of equations determines the scalar $\cF$ and the one-form $\omega$. 

In the class of superstratum solutions that we will consider, $Z_2$ has the simple form
\be
Z_2 \;=\; \frac{Q_5}{\Sigma}  \,.
\ee
The first important feature of the solutions is encoded in the function $Z_4$ which enters directly into the Type IIB NS-NS two-form $B_2$, and the RR forms $C^{(0)}$ and $C^{(4)}$, and also into the metric via the combination $\cP=Z_1Z_2-Z_4^2$.
The function $Z_4$ takes the general form (more generally a phase could also be introduced in the definition of $Z_4$)
\be
\label{Z4_solngen}
Z_4 \;=\; R_y\sum\limits_{k,m,n,q} \delta_{q,0} \, b_{4}^{k,m,n,q} \, {\Delta_{k,m,n}\over \Sigma} \cos\hat{v}_{k,m,n}\,,
\ee
where $b_{4}^{k,m,n,q}$ are real coefficients (the inclusion of $q$ in the indices is somewhat superfluous because of the $\delta_{q,0}$, however we choose to keep the notation general), and where
\begin{align} 
\begin{aligned}
  \Delta_{k,m,n} &~\equiv~
 \left(\frac{a}{\sqrt{r^2+a^2}}\right)^k
 \left(\frac{r}{\sqrt{r^2+a^2}}\right)^n 
 \cos^{m}\theta \, \sin^{k-m}\theta \,, \cr
 \hat{v}_{k,m,n} &~\equiv~  (m+n) \frac{\sqrt{2}\,v}{R_y} + (k-m)\phi - m\psi \,,
\qquad \Sigma \,\equiv\, r^2 + a^2 \cos^2\theta\,.
\end{aligned} 
\label{Delta_v_kmn_def}
\end{align}

The ansatz for $Z_1$ involves a linear combination of terms similar to those appearing in $Z_4$, with coefficients chosen to facilitate the construction of smooth solutions without horizons. This procedure is known as ``coiffuring''~\cite{Mathur:2013nja,Bena:2013ora,Bena:2015bea}. In practical terms, this means making the combination $\cP=Z_1Z_2-Z_4^2$ have desired properties, which in the simplest cases means arranging that $\cP$ is independent of $\hat{v}_{k,m,n}$. Several families of asymptotically AdS$_3$ solutions have this property, and in fact have the property that the full metric is also independent of the phase $\hat{v}_{k,m,n}$ and all explicit dependence on this phase is in the matter fields. We will discuss the explicit form of $Z_1$ that exhibits ``coiffuring'' once we specialize the discussion to the solutions that we consider in this paper.

The proposed CFT interpretation of the superstratum solutions involves coherent superpositions of several strands of the following type. The states are labelled by integers $(m,n,k,q)$ with\footnote{We use the notation of~\cite{Heidmann:2019zws} which differs from that of CRS~\cite{Ceplak:2018pws} by $(m-q)_{\mathrm{here}} = m_{\mathrm{CRS}}$ and $(n-q)_{\mathrm{here}} = n_{\mathrm{CRS}}$.} $q=0,1$; $n \ge 1$; and $k >0, k -q \ge m \ge 1$. For ease of notation it is convenient to define the states in the NS-NS sector, where they are given by~\cite{Bena:2015bea,Bena:2016agb,Bena:2016ypk,Bena:2017xbt,Ceplak:2018pws}
\bea \label{eq:cft-states}
\!\! |k,m,n,q\rangle^{\NS} &\!\!=\!\!& \frac{1}{(m-q)!(n-q)!}
  (J^+_0)^{m-q} L_{-1}^{n-q}  \left(G_{-\frac12}^{+1}G_{-\frac12}^{+2} + \frac{1}{k} J^+_0 L_{-1}\right)^q  |O^{--}\rangle_k^{\NS} \,,  
\eea
with $ |O^{--}\rangle_k^{\NS}$ the NS-sector anti-chiral primary corresponding to the RR ground state $\ket{00}_k$.
Then the states we are interested in are the RR states obtained by performing left and right spectral flow transformations with parameters $(1/2, 1/2)$, and for ease of notation we shall denote the resulting RR states by $\ket{k_i,m_i,n_i,q_i}$, where $i$ runs over the different types of superstratum strands that are present in a given state. Our spectral flow conventions are recorded in Eqs.\;\eq{spectral flow transformation}--\eq{eq:sf-states} and are such that spectral flow with parameters $(1/2, 1/2)$ on an individual copy of the CFT maps the NS-NS vacuum to the RR ground state $\ket{++}$.

We are interested in coherent superpositions of the states involving $N_i$ copies of the above superstratum-type strands $\ket{k_i,m_i,n_i,q_i}$ and $N^{(s)}_k$ copies of the bosonic RR ground state strands $|s\rangle_k$ introduced around Eq.\;\eq{eq:RRstate}:
\be
\psi_{\{N^{(s)}_k, n_i \}}\equiv \prod_{s=1}^4 \prod_{k} |s\rangle_k^{N^{(s)}_k} \prod_i  \ket{k_i,m_i,n_i,q_i }^{N_i}\,.
\ee
The resulting family of (non-normalized) CFT states $\psi(\{A^{(s)}_k,B_{i}\})$ is defined, in a way similar to \eqref{eq:coherent}, as 
\be\label{eq:coherentsuperstrata}
\psi(\{A^{(s)}_k,B_{i}\}) \;\equiv\; {\sum_{\{N_i,N^{(s)}_{k}\}}}^{\!\!\!\!\!\!\prime} \,\,
\left[\;\prod_{s=1}^4 \prod_{k}  \Big(A^{(s)}_k\,|s\rangle_k\Big)^{N^{(s)}_k } \prod_i \Big(B_i \ket{k_i,m_i,n_i,q_i }\Big)^{N_i} 
\right]\, .
\ee
where the prime on the overall sum indicates that it is a restricted sum (as in Eq.\;\eq{eq:coherent}) over all states whose total number of copies adds up to $N$:
\begin{equation}\label{eq:windingconstraint-SS}
\sum_{k,s} k N_k^{(s)}+ \sum_{i}k_i N_i \,=\, N\,.
\end{equation}
Having defined the general class of superstratum states, we now specialize to those that we will consider in this paper. We consider states with one type of ground state strands, with winding $k=1$ and polarization $s=++$, and one type of superstratum strand:
\be
\psi(A_1,B_{k, m, n, q}) \;=\; \sum_{p=1}^{N/p} \,
\Big(A_1 \ket{++}_1\Big)^{N-pk} \Big(B_{k,m,n,q}\, \ket{k,m,n,q}\Big)^{p} \,.
\ee
This class is both sufficiently tractable and sufficiently interesting to enable the new precision holographic tests that follow.

The computations in the following subsections make use of a number of technical results, such as the norm of the states $\psi_{\{N_1\,,N_{k,m,n,q} \}}$, the average numbers $\overline{N}_i$ of strands in the coherent state \eqref{eq:coherentsuperstrata}, and the map between the CFT parameters $A_1$, $B_{k,m,n,q}$ and the coefficients $a$, $b_4^{k,m,n,q}$ that define the supergravity solution. For the examples considered below, it will be sufficient to present these results for $q=0$, whose derivation can be found in \cite{Bena:2017xbt}:
\be\label{eq:normsuperstrata}
\left | \psi_{\{N_1,\,N_{k,m,n,0} \}} \right |^2 \,=\, \frac{N!}{N_1!}\prod_{k,m,n} \frac{1}{N_{k,m,n,0}!} \left[ \frac{1}{k}\,\binom{k}{m} \binom{n+k-1}{n}\right]^{N_{k,m,n,0}},
\ee
  \begin{equation}
  \label{eq:averagesuperstrata}
  \overline{N}_1 \,=\, |A_1|^2\;,
  \qquad~~~
  k \overline{N}_{k,m,n,0} \,=\, {{k}\choose{m}} {{n+k-1}\choose{n}} |B_{k,m,n,0}|^2\;,
\end{equation}
\be
\label{eq:CFTgravsuperstrata}
|A_1| \,=\, R_y\sqrt{\frac{N}{Q_1Q_5}}\, a
\;,\qquad~~
|B_{k,m,n,0}| \,=\, R_y\sqrt{\frac{N}{2Q_1Q_5}}\, {{k}\choose{m}}^{-1} {{n+k-1}\choose{n}}^{-1}  b_4^{k,m,n,0} ~.
\ee

\subsection{Holographic tests of superstrata with the operator $O_2$}

We now make the first precision holographic test of superstrata at dimension two, focusing on the expectation value of the operator $O_2$. Since the one-point function of $O_2$ is extracted from the metric function $Z_4$, which is the basic ingredient in the construction of the superstrata solutions, these are the most direct tests of the identification between superstrata and CFT states.

\subsubsection*{Superstrata with $k=2$, $m=1$}

We now consider the following set of states:
\begin{equation}\label{O2 m1 state-pre}	
\sum\limits_{p=1}^{\frac{N}{2}}\big(A \ket{++}_1\big)^{N-2p} \left(B \frac{(L_{-1}-J^3_{-1})^n}{n!} J^+_{-1}\ket{00}_2\right)^{p} \,.
\end{equation}
To begin with we will set $n=0$, before extending to general $n$. We thus first consider the states
\begin{equation}\label{O2 m1 state}	
\sum\limits_{p=1}^{\frac{N}{2}}\big(A \ket{++}_1\big)^{N-2p} \big(B J^+_{-1}\ket{00}_2\big)^{p} \,.
\end{equation}
In the CFT, of the operators $O_2$ and $(\Sigma_2 \cdot O)$ entering in the holographic dictionary \eqref{eq:mapdim2O2}, only the single-trace $O_2$ has a non trivial expectation value: the expectation value of the operators $O$ and $\Sigma_2$ are zero on this state, thus also that of the double-trace $(\Sigma_2 \cdot O)$ is zero. 

Moreover, since the strands $\ket{++}_1$ and $J^+_{-1}\ket{00}_2$ carry spin $(\frac{1}{2},\frac{1}{2})$ and $(1,0)$ respectively, by angular momentum conservation we conclude that only $O_2^{0-}$ and its hermitian conjugate have non-vanishing one-point functions. The basic process is that in which $O_2^{0-}$ links two strands $\ket{++}_1$ into a strand $J^+_{-1}\ket{00}_2$ and the corresponding amplitude is 
	\begin{equation}\label{eq:onepointO20m}	
	{}_{(12)}\bra{00}J^-_{+1}O_2^{0-}(v,u)\ket{++}_{(1)}\ket{++}_{(2)}\;=\;\sqrt{2}e^{i\frac{\sqrt{2}\,v}{R_y}}\,.
	\end{equation}
In deriving this result we have used the fact that the ground state is annihilated by the positive modes of the current operator to replace $J^-_{+1}O_2^{0-}(v,u)$ by their commutator\footnote{The factor $\sqrt{2}$ in the commutator \eqref{eq:communatorJplus} ensures, as usual, that all components of $O^{a\dot a}_2$ have unit norm.}
\begin{equation}\label{eq:communatorJplus}
[J^-_{+1},O_2^{0-}(v,u)]\;=\;\sqrt{2}\,e^{i\frac{\sqrt{2}\,v}{R_y}}\,O_2^{--}(v,u)\,,
\end{equation}
and the hermitian conjugate of the second relation in \eqref{eq:defO12}. Note that it is important to insert the operator $O_2^{0-}$ at a generic worldsheet point $(v,u)$ to obtain a non-trivial result: had we inserted it at past infinity, it would have killed the initial state $\ket{++}_{(1)}\ket{++}_{(2)}$. 

We must now dress the result \eqref{eq:onepointO20m} with the proper combinatorial factor: the operator $O_2^{0-}$ can act on any of the ${N-2p}\choose{2}$ pairs of $\ket{++}_1$ to produce the state $J_{-1}^+\ket{00}_2$. Using \eqref{eq:normsuperstrata} and requiring that both sides of the equation contain the same number of terms, we obtain
	\begin{equation}\label{O2 m1 process}
	O^{0-}_2\big(\ket{++}_1\big)^{N-2p} \big(J_{-1}^+\ket{00}_2\big)^{p} \;=\;\frac{p+1}{\sqrt{2}}e^{i\frac{\sqrt{2}\:\! v}{R_y}}\big(\ket{++}_1\big)^{N-2(p+1)} \big(J_{-1}^+\ket{00}_2\big)^{p+1} \,.
	\end{equation}
This implies that
\begin{equation}\label{O2 m1 vev}\begin{split}
\big\langle O_2^{0-}(v,u) \big\rangle \,=\, \frac{A^2}   {\sqrt{2}B}\,\bar{p}\,e^{i\frac{\sqrt{2}\:\! v}{R_y}} \,=\, \frac{A^2 \bar{B}}{\sqrt{2}}e^{i\frac{\sqrt{2}\:\! v}{R_y}} 
\,=\, \frac{N^{\frac{3}{2}} \,R_y^3}{4(Q_1Q_5)^{\frac{3}{2}}}a^2\,\bar{b}\,e^{i\frac{\sqrt{2}\:\! v}{R_y}} \,,
\end{split}
\end{equation}
where we have used \eqref{eq:averagesuperstrata} to compute $\bar p$ and \eqref{eq:CFTgravsuperstrata} to express the final result in terms of the gravity parameters. 

On the supergravity side, we require the first non-trivial terms in the large $r$ expansion of the function $Z_4$ given in \eqref{Z4_solngen} where $k=2$, $m=1$, $n=0$, $q=0$ and $b_4^{2,1,0,0}= b\,$: 
\begin{equation}\label{O2 m1 geom exp}	\begin{split}
Z_4\;\sim\;\frac{\sqrt{Q_1Q_5}}{r^4} \frac{R_y \,a^2 \,b}{2 \sqrt{6}\sqrt{Q_1Q_5}}
\left(-e^{i\frac{\sqrt{2}\:\! v}{R_y}}Y_2^{0,1}+e^{-i\frac{\sqrt{2}\:\! v}{R_y}}Y_2^{0,-1}\right)\, .
\end{split}
\end{equation}
Comparing the result~\eqref{O2 m1 vev} with~\eqref{O2 m1 geom exp} using the dictionary in~\eqref{eq:mapdim2O2final} and~\eqref{eq:deftildeO2}, one obtains exact agreement.

It is now straightforward to generalize the $n=0$ computation to the general set of states \eq{O2 m1 state-pre}.
On the CFT side the computation proceeds along the same lines as before, with the only difference that the correlator \eqref{eq:onepointO20m} should be replaced by
\begin{equation}\label{eq:onepointO20mn}	
	{}_{(12)}\bra{00}\frac{(L_{1}-J^3_{1})^n}{n!} J^-_{+1}O_2^{0-}(v,u)\ket{++}_{(1)}\ket{++}_{(2)}\;=\;\sqrt{2}e^{i\:\! (n+1)\frac{\sqrt{2}\:\! v}{R_y}}\,.
	\end{equation}
The extra factor $e^{i \:\! n\frac{\sqrt{2}\:\! v}{R_y}}$ is produced by commuting the operator $(L_{1}-J^3_{1})^n$ with $O^{--}_2(v,u)$, using
\begin{equation} \label{eq:gen-n}
\left[(L_{1}-J^3_{1})^n, O^{--}_2(v,u)\right] \;=\; n!\,e^{i \:\! n\frac{\sqrt{2}\:\! v}{R_y}}\,O^{--}_{2,0}\,,
\end{equation}
where $O^{--}_{2,0}$ denotes the zero-mode of $O^{--}_2$, which is the only one contributing to the correlator after having eliminated the momentum-carrying operators. On the gravity side, it follows immediately from \eqref{Z4_solngen} that the only modification to $Z_4$ at order $r^{-4}$ is an extra factor $e^{i\:\! n\frac{\sqrt{2}\:\! v}{R_y}}$. We thus see that the exact agreement persists for any value of $n$.

\subsubsection*{Superstrata with $k=2$, $m=2$}

As a further consistency check, we consider the set of superstratum states with $k=2$, $m=2$:
\begin{equation}\label{O2 m2 state-pre}	
\sum\limits_{p=1}^{\frac{N}{2}}(A \ket{++}_1)^{N-2p} \left( B\,\frac{(L_{-1}-J^3_{-1})^n}{n!}\frac{(J^+_{-1})^2}{2}\ket{00}_2\right)^{p} \,.
\end{equation}
We follow the same presentation and first set $n=0$, before extending to general $n$.
Thus, we first consider the coherent state
\begin{equation}\label{O2 m2 state}	\begin{split}
\sum\limits_{p=1}^{\frac{N}{2}}(A \ket{++}_1)^{N-2p} \left( B\,\frac{(J^+_{-1})^2}{2}\ket{00}_2\right)^{p} .
\end{split}
\end{equation}
The strand $(J^+_{-1})^2\ket{00}_2$ carries spin $(2,0)$, thus, by conservation of angular momentum, we conclude that only the operator $O^{+-}_2 =\frac{1}{2}\left[(\tilde{J}^-_0)^2,O_2^{++}\right]$ and its hermitian conjugate will have non-trivial expectation values; the expectation value of the multi-trace $(\Sigma_2\cdot O)$ is trivially zero. This operator carries out the fundamental process
\begin{equation}
O^{+-}_2(v,u)\,\ket{++}_1\ket{++}_1=e^{i\frac{2\sqrt{2}v}{R_y}}\frac{(J^+_{-1})^2}{2}\ket{00}_2\,,
\end{equation}
where we have used the commutation relation $[(J^-_1)^2,O^{+-}_2(v,u)]=2\,e^{i\frac{2\sqrt{2}v}{R_y}} \,O^{--}_2(v,u)$, the relation defining $O^{--}$, given by the hermitian conjugate of \eqref{eq:defO12}, and the fact that $\frac{(J^+_{-1})^2}{2}\ket{00}_2$ has unit norm.
The complete action of the operator $O_2^{+-}$ on the state is obtained implementing the appropriate combinatorial factor (which follows, as usual, noticing that $O_2^{+-}$ can choose among ${N-2p}\choose{2}$ pairs of $\ket{++}_1$ and imposing that the norms on the two sides of the equation are equal). We obtain
\begin{equation}	
O_2^{+-}(v,u)\left[( \ket{++}_1)^{N-2p} \left(\frac{(J^+_{-1})^2}{2}\ket{00}_2\right)^{p}\,\right]\;=\;
e^{i\frac{2\sqrt{2}v}{R_y}}(p+1)(\ket{++}_1)^{N-2p-2} \left(\frac{(J^+_{-1})^2}{2}\ket{00}_2\right)^{p+1} .
\end{equation}
This gives rise to the expectation value
\begin{equation}\label{O2 m2 vev}	
\big\langle O_2^{+,-}(v,u) \big\rangle \,=\, e^{i\frac{2\sqrt{2}v}{R_y}}\bar{p} \frac{A^2}{B}
\,=\,e^{i\frac{2\sqrt{2}v}{R_y}} \frac{A^2\bar{B}}{2}
\,=\,e^{i\frac{2\sqrt{2}v}{R_y}} 
\frac{N^{\frac{3}{2}}\,R_y^3}{2\sqrt{2}(Q_1Q_5)^{\frac{3}{2}}}a^2\,\bar{b}\,.
\end{equation}

Expanding the $Z_4$ function of the dual geometry \eqref{Z4_solngen} (with $k=2$, $m=2$, $n=q=0$, $b_4^{2,2,0,0}= b$) for large $r$ up to the first non-trivial order, we obtain
\begin{equation}\label{O2 m2 geometry coeff}	\begin{split}
Z_4~\sim~\frac{\sqrt{Q_1Q_5}}{r^4} \frac{R_y \,a^2 b}{2 \sqrt{3}\sqrt{Q_1Q_5}}
\left(e^{i\frac{2\sqrt{2}v}{R_y}}Y_2^{-1,1}+e^{-i\frac{2\sqrt{2}v}{R_y}}Y_2^{+1,-1}\right) .
\end{split}
\end{equation}
Eqs.\;\eqref{O2 m2 vev} and~\eqref{O2 m2 geometry coeff} are in exact agreement with the dictionary given in Eqs.\;\eqref{eq:mapdim2O2final} and~\eqref{eq:deftildeO2}. 

As explained around Eq.\;\eq{eq:gen-n}, it is straightforward to extend this result to the states with general $n$ given in Eq.\;\eq{O2 m2 state-pre}: both the CFT and the gravity results are simply multiplied by the factor $e^{i\:\! n\frac{\sqrt{2}\:\! v}{R_y}}$.

\subsection{Holographic tests of superstrata with the operators $\Omega^{00}$ and $\Sigma_3^{00}$}
\label{sec:coiffuring}
We now consider the class of states with $m=1$, $n=0$, $q=0$ and general (positive integer) $k$:
\begin{equation}
\sum_{p=1}^{N/k} \big(A\,\ket{++}_1\big)^{N-k\:\! p} \big(B\,J^+_{-1}\ket{00}_k\big)^p\,.
\end{equation}
The one-point function of $O_2$ in the state with $k=2$ has already been considered in the previous subsection; here we concentrate on the other dimension-two operators, with the purpose of testing the dictionary \eqref{eq:mapdim2bis}. This enables us to check some features of the dual geometry other than $Z_4$, and in particular the metric function $Z_1$. Setting for ease of notation $b_4^{k,1,0,0}=b$, for this class of metrics $Z_1$ is given by (see e.g.~\cite[Eq.\;(4.3)]{Bena:2017xbt}):
\be\label{eq:Z1superstrata}
Z_1 \;=\; \frac{Q_1}{\Sigma}+ \frac{R_y^2 \,b^2}{2 Q_5}\,\frac{\Delta_{k,1,0}}{\Sigma}\,\cos\hat v_{2k, 2,0}\,,
\ee
where
\be\label{eq:Q1superstrata}
Q_1 \;=\; \frac{R_y^2}{Q_5} \left(a^2 + \frac{b^2}{2\:\!k} \right) .
\ee
The term proportional to $Q_1$ is the standard term encoding the dependence on the D1 charge; the term proportional to $b^2$ is more subtle, since it cannot be inferred simply on the basis of the global charges or of the supergravity equations, which would be satisfied also in the absence of that term. Its presence is however crucial for the smoothness of the solution.
The general mechanism by which regularity is ensured in the superstratum construction has been dubbed ``coiffuring"~\cite{Mathur:2013nja,Bena:2013ora}, and in this example it amounts to choosing the ansatz for the function $Z_1$ such that the combination $\mathcal{P}=Z_1 Z_2 - Z_4^2$ is independent of $v$. 

The holographic dictionary can provide a more direct CFT understanding of the coiffuring construction: we will show that the $b^2$ contribution to $Z_1$ originates from the mixing of both $\Omega$ and $\Sigma_3$ with the double-trace operator $(O\cdot O)$.  

We first consider the second line of the holographic dictionary \eqref{eq:mapdim2bis}, which involves the expectation values of $\Omega^{00}$ and $(J\cdot \tilde J)^{00}$ (in these states the one-point function of $(\Sigma_2 \cdot \Sigma_2)^{00}$ is trivially zero for any $k$). Since these operators act in a way that has essentially already been explained in Section~\ref{sec:tests}, we will be brief in the following. The operator $\Omega^{00}$ acts non-trivially only on the $\ket{++}_1$ strands, for which $ \Omega^{00} \ket{++}_1 = 1/2 \,\ket{++}_1$, so we obtain
\begin{equation}
\big\langle \Omega^{00} \big\rangle \,=\, \frac{|A|^2}{2} \,=\, \frac{1}{2}\,\frac{N\,R_y^2}{Q_1 Q_5}\,a^2 \,=\, \frac{1}{2}\,\frac{N\,R_y^4}{(Q_1 Q_5)^2}\,a^2 \left(a^2 + \frac{b^2}{2k} \right),
\end{equation}
where we used \eqref{eq:CFTgravsuperstrata} and, for later convenience, the regularity constraint \eqref{eq:Q1superstrata}. The expectation value of the double-trace $(J\cdot \tilde J)^{00}$ can be expressed, as usual, as the product of the expectation values of $J^3$ and $\tilde J^3$: 
\begin{equation}
\big\langle (J\cdot \tilde J)^{00} \big\rangle \,=\,\frac{2}{N} \,\big\langle J^3 \rangle\, \langle \tilde J^3 \rangle \,=\,\frac{|A|^2}{N}\left(\frac{|A|^2}{2} +|B|^2\right)\,=\,\frac{1}{2} \,\frac{N\,R_y^4}{(Q_1 Q_5)^2}\,a^2\left(a^2+\frac{b^2}{k^2}\right).
\end{equation}
Substituting in the second line of \eqref{eq:mapdim2bis}, we find a value of $\tilde h_{(0,0)}$ in exact agreement with the one extracted from the geometry: 
\begin{equation}
\tilde h_{(0,0)} \;=\; \frac{N^{1/2}\,R_y^4}{(Q_1 Q_5)^2}\,\frac{k-2}{4\,k^2}\,a^2 b^2\,.
\end{equation}

The first line of the holographic dictionary \eqref{eq:mapdim2bis} works in a more interesting way, and it requires us to distinguish the states with $k=1$ from the ones with $k>1$. For $k=1$ we have $\langle \Sigma_3 \rangle =0$, however the following components of the double-trace $(O\cdot O)$ play a role:
\begin{equation}
(O\cdot O)^{00} = \frac{1}{N}\sum_{r\neq s} \big(O^{++}_{(r)} O^{--}_{(s)} + O^{+-}_{(r)}O^{-+}_{(s)}\big)~,\qquad\quad (O\cdot O)^{+-} = \frac{1}{N}\sum_{r\neq s} O^{+-}_{(r)} O^{+-}_{(s)}\,,
\end{equation}
as well as the hermitian conjugate $(O\cdot O)^{-+}$. On the CFT side the expectation values are straightforward to  compute as the product of the expectation values of the single-particle operators $O^{+-}$ and $O^{-+}$, which were derived in Eqs.\;(4.38), (4.39) of \cite{Giusto:2015dfa}:
\begin{equation}
\big\langle (O\cdot O)^{00} \big\rangle \;=\; -\frac{|A|^2\,|B|^2}{N}~,\qquad\quad \big\langle (O\cdot O)^{+-} \big\rangle \;=\; e^{i \frac{2\sqrt{2}\:\! v}{R_y}}\,\frac{A^2\,\bar B^2}{N}\,.
\end{equation}
On the gravity side the term responsible for $\langle (O\cdot O)^{+-} \rangle$ is the term quadratic in $b$ in the metric function $Z_1$ \eqref{eq:Z1superstrata}, from which one extracts
\begin{equation}
h_{(-,+)}\,=\,(h_{(+,-)})^* \,=\, \frac{N^{1/2}R_y^4}{Q_1^2Q_5^2}e^{i\frac{2\sqrt{2}v}{R_y}}\frac{a^2b^2}{4\sqrt{3}}\qquad \mathrm{and}\qquad h_{(0,0)}\,=\,-\frac{N^{1/2}R_y^4}{Q_1^2Q_5^2}\frac{a^2b^2}{4\sqrt{3}}
\,,
\end{equation}
which agree precisely with the CFT results. As we discussed below Eq.~\eqref{eq:Z1superstrata}, the term contributing to $h_{(-,+)}$ is the one deduced, quite indirectly, from the ``coiffuring'' method. It is satisfying to see that holography provides a sharp CFT explanation of this supergravity construction. 

When $k>1$ the relevant operator is $\Sigma_3^{00}$, which, as we have already seen, has to be analyzed with some care. The non-trivial part of the computation is in the derivation of the coefficient $C_{k3k}^{00(m=1)}$, which captures the action of the twist-three operator on a particular pair of states $\ket{++}_1$ and $J^+_{-1}\ket{00}_k$:
\begin{equation}
\sigma^{00}_{(3)}\,\Big(\ket{++}_1\,J^+_{-1}\ket{00}_k\Big) \;=\; C_{k3k}^{00(m=1)}\,\Big(\ket{++}_1\,J^+_{-1}\ket{00}_k\Big) \,.
\end{equation}
Similarly to our explanation of the process \eqref{eq:sigma3process}, the twist operator $\sigma^{00}_{(3)}$ can cut the strand $\ket{00}_k$ and join it with the strand $\ket{++}_1$, while at the same permuting the spins of the two copies involved in the process. To our knowledge the coefficient $C_{k3k}^{00(m=1)}$ does not appear in the literature, and we thus derive it in Appendix~\ref{sec:Sigma3app}, by evaluating the three-point function \eqref{correlator R sector ++1J-00k}. The result is
\begin{equation} \label{eq:use-app-2}
C_{k3k}^{00(m=1)} \;=\; \frac{k-2}{6 k}\,.
\end{equation}
When acting on the full state, the twist operator $\Sigma_3^{00}$ can act on any of the $N-pk$ strands $\ket{++}_1$ and on any of the $p$ strands $J^+_{-1}\ket{00}_k$, and can cut the latter in $k$ positions (after this choice is made, the permutation by which the twist operator can act is completely fixed); this translates, according to the usual logic, into the identity
\begin{equation}
\Sigma_3^{00}\,\left(\ket{++}_1^{N-k p}\,\big(J^+_{-1}\ket{00}_k\big)^p \right)\;=\;C_{k3k}^{00(m=1)}\,(N-k p)\,p\,k\, \ket{++}_1^{N-k p}\,\big(J^+_{-1}\ket{00}_k\big)^p\,,
\end{equation}
and thus, using the result~\eqref{eq:averagesuperstrata} to compute $\bar p$ and the CFT--gravity parameter map \eqref{eq:CFTgravsuperstrata}, one arrives at 
\begin{equation}
\big\langle \Sigma_3^{00} \big\rangle \,=\, C_{k3k}^{00(m=1)}\,k\,A^2\,B^2 \,=\, \frac{k-2}{12\, k^2}\,\frac{N^2\,R_y^4}{(Q_1 Q_5)^2}\,a^2\,b^2\,.
\end{equation}
 From the dual gravity solution one extracts:
 \begin{equation}
 h_{0,0} \;=\;  \frac{k-2}{4\sqrt{3}\, k^2}\,\frac{N^{1/2}\,R_y^4}{(Q_1 Q_5)^2}\,a^2\,b^2\,,
 \end{equation}
 which agrees precisely with the prediction of the map \eqref{eq:mapdim2bis}.

\subsection{A holographic test of supercharged superstrata}

A more recently constructed, and therefore less-studied, class of superstrata is that of \cite{Ceplak:2018pws}, where some of the momentum is carried by the CFT supercurrents $G$. We will focus here on the simplest state in that class, the one with $k=2$, $m=1$, $n=0$, $q=1$ in the notation of \eqref{eq:cft-states}, which we rewrite in the Ramond sector as:
\begin{equation}\label{eq:CRSstate}
\sum_{p=1}^{N/2} (A \ket{++}_1)^{N-2p} \left[ B \left(G^{+1}_{-1} G^{+2}_{-1}+\frac{1}{2} J^+_{-1}(L_{-1}-J^3_{-1})\right)\ket{00}_2\right]^p .
\end{equation}
We now verify the important feature of the supergravity solution dual to this state, namely that $Z_4=0$ (as indicated by the $\delta_{q,0}$ in Eq.\;\eqref{Z4_solngen}). For consistency with the holographic dictionary (\ref{eq:mapdim2O2final})--(\ref{eq:deftildeO2}), one expects that the expectation values of $O_2$ and $(\Sigma_2\cdot O)$ vanish. While this is obvious for the double-trace $(\Sigma_2\cdot O)$, at first sight one could have a non-vanishing expectation value for $O_2^{0+}$, generated by the correlator
\begin{equation}
{}_1\bra{++} \,{}_1\bra{++} \,O_2^{0+}(z,\bar z) \left(G^{+1}_{-1} G^{+2}_{-1}+\frac{1}{2} J^+_{-1}(L_{-1}-J^3_{-1})\right)\ket{00}_2\,.
\end{equation}
It is simpler to perform the computation in the NS sector, where this correlator becomes
 \begin{equation}\label{eq:NScorrelatorO2}
{\bar z} \;{}_{\NS}\bra{0} \,O_2^{0+}(z,\bar z) \left(G^{+1}_{-1/2} G^{+2}_{-1/2}+\frac{1}{2} J^+_{0} L_{-1}\right) \ket{O^{--}}_{2}^{\NS}\,,
\end{equation}
with ${}_{\NS}\bra{0}$ the NS vacuum and $\ket{O^{--}}_{2}^{\NS}\equiv O^{--}_2(0,0) \ket{0}^{\NS}$ the anti-chiral-primary state with $h=\bar h=1$ and $j=\bar j=-1$ introduced after Eq.~\eqref{eq:cft-states}. One can write
\begin{equation}\label{eq:L1expansion}
O_2^{0+}(z,\bar z) \;=\; [J_0^-, O_2^{++}(z,\bar z)]\;=\;[J_0^-, O_2^{++}(\infty)]+ z^{-1}\,[J_0^-, [L_{1},[O_2^{++}(\infty)]]+\dots\,,
\end{equation}
where the dots represent terms with higher powers of $L_1$ or $\tilde L_1$, which cannot contribute to the correlator. Inserting \eqref{eq:L1expansion} in \eqref{eq:NScorrelatorO2}, one finds that the correlator is proportional to 
\begin{equation}
{}^{\NS}_{\,\,\,\,2}\bra{O^{--}} \,J^-_0 \,L_{1}\left(G^{+1}_{-1/2} G^{+2}_{-1/2}+\frac{1}{2} J^+_{0} L_{-1}\right) \ket{O^{--}}_{2}^{\NS}=\,{}^{\NS}_{\,\,\,\,2}\bra{O^{--}}  \,J^-_0 \left(-J_0^+ + J^+_{0} L_{0}\right) \ket{O^{--}}_{2}^{\NS}\:\!=\:\!0\,,
\end{equation}
where we have used the chiral algebra commutation relations and the fact that $L_0 \ket{O^{--}}_{2}^{\NS}=\ket{O^{--}}_{2}^{\NS}$, as in \cite[Eq.\;(2.7)]{Ceplak:2018pws}. The vanishing of $Z_4$ for the state \eqref{eq:CRSstate} is thus in exact agreement with the CFT prediction.

\section{Discussion}
\label{sec:disc}

The main result of this article is the derivation of the holographic map relating the expectation values of chiral primary operators (CPOs) of dimension $(1,1)$ in a 1/4 or 1/8-BPS state of the D1-D5 CFT with the geometric coefficients extracted from the asymptotic expansion of the supergravity solution dual to the state. The result, which is valid for the class of $\mathcal{M}$-invariant supergravity solutions with a flat four-dimensional base space, and which includes all possible mixings between single-trace and double-trace operators, is summarized in Eqs.\;\eqref{eq:mapdim2bis}--\eqref{eq:deftildeO2-1}.

This holographic dictionary should be useful for understanding the nature of black hole microstates and, in particular, for the development of the fuzzball program, in several ways. Given a smooth horizonless geometry carrying the right charges (D1, D5 and P charges in our duality frame) its identification with a state of the CFT dual to the black hole is in most instances a difficult task, mainly because the point in the CFT moduli space where one can easily describe the states is far from the point where the classical gravity description holds. As mentioned in the Introduction, the expectation values of CPOs in 1/4 and 1/8-BPS states are protected quantities~\cite{Baggio:2012rr} that provide a useful bridge between the two descriptions. We have tested the correspondence between supergravity solutions and the proposed dual CFT states both for several 1/4-BPS RR ground states, which had already been extensively studied from many other angles, but also for some of the most recently constructed and less-studied 1/8-BPS D1-D5-P states, including a set of ``superstratum'' solutions. The precise match between CFT and gravity predictions, which is found in all cases, required the knowledge of correlation functions between operators of the orbifold CFT in non-trivial twist sectors -- some already known in the literature, and others that we have computed in this work. The high level of non-triviality of the agreement appears to show, beyond reasonable doubt, that the geometries of the superstrata constructed in \cite{Bena:2015bea,Bena:2016agb,Bena:2016ypk,Bena:2017geu,Bena:2017upb,Bena:2017xbt,Bena:2018bbd,Bakhshaei:2018vux,Bena:2018mpb,Ceplak:2018pws,Heidmann:2019zws} are indeed dual to the proposed family of microstates of the CFT that has been used to derive the entropy of the D1-D5-P black hole. 

The holographic point of view also shows how some features of the gravitational solution that were determined by requiring the regularity of the geometry could in principle have been predicted simply by computing correlators in the orbifold CFT: the D1-D5-P state examined in Section~\ref{sec:coiffuring}, for instance, shows how the so-called ``coiffuring" technique~\cite{Mathur:2013nja,Bena:2013ora} used in the superstratum construction is a reflection of the mixing between the single-trace and double-trace operators. Related observations were made in \cite{Giusto:2015dfa}. A more general coiffuring has been recently employed in \cite{Heidmann:2019zws} to cancel the singularities in a family of multi-mode three-charge superstrata, and it would be interesting to investigate whether there exists a similar dual CFT understanding of this construction. Work in this direction is in progress. More generally, this point of view could also prove useful in finding new microstate solutions, as in many cases computing free CFT correlators is easier than solving the non-linear supergravity equations. 

To conclude, holography has been an essential tool in the development of the fuzzball program, and our results provide a new sharper formulation of the holographic dictionary. We expect that our results should prove useful to clarify both the power and the limits of supergravity to describe microstates of black holes.

\vspace{5mm}
\section*{Acknowledgements}

We thank Joan Garcia i Tormo, Rodolfo Russo, Michele Santagata, Kostas Skenderis and Marika Taylor for fruitful discussions.
This work was supported in part by the MIUR-PRIN contract 2017CC72MK003.
The work of SR was supported by a Royal Society URF Enhancement Award.
The work of DT was supported by a Royal Society Tata University Research Fellowship.

\newpage

\appendix

\section{Spherical harmonics on S$^3$}
\label{sec:sphericalharm}
The spherical harmonics on S$^3$ are a representation of the isometry group of the three-sphere $SO(4)\simeq SU(2)_L\times SU(2)_R$. 
We will use spherical coordinates in the $\mathbb{R}^4$ base space that are related to the  Cartesian coordinates via 
\begin{equation}\label{cartesian to spherical coordinates} \begin{split}
x^1&=r\sin\theta\cos\phi \;,\qquad x^2=r\sin\theta\sin\phi\,,\\
x^3&=r\cos\theta\cos\psi \;,\qquad x^4=r\cos\theta\sin\psi\,,
\end{split}
\end{equation}
where $\theta\in[0,\frac{\pi}{2}]$ and $\psi,\phi\in[0,2\pi)$. With this coordinate choice, the S$^3$ line element $ds_3^2$ is given by $ds_3^2=d\theta^2+\sin^2\theta d\phi^2+\cos^2\theta d\psi^2 $.
We will denote a degree $k$ scalar harmonic with $Y_k^{m,\tilde{m}}$, where $(m,\tilde{m})$ are the spin charges under $(J^3,\tilde{J}^3)$. 
Denoting the volume of S$^3$ by $\Omega_3=2\pi^2$, we use normalized spherical harmonics
\begin{equation}\label{normalization sph harmonic} \begin{split}
\int Y_{k_1}^{*m_1,\tilde{m}_1}Y_{k_2}^{m_2,\tilde{m}_2}\;=\;\Omega_3 \,\delta_{k_1,k_2}\delta^{m_1,m_1}\delta^{\tilde{m}_1,\tilde{m}_2}\,.
\end{split}
\end{equation}
The generators of the isometry group of S$^3$, written in terms of the standard $SU(2)$ generators, are
\begin{equation}\label{generators isometry S3} \begin{split}
	J^{\pm}&=\frac{1}{2}  e^{\pm i(\phi+\psi)}  (\pm \partial_\theta +i \cot\theta  \partial_\phi  - i \tan\theta  \partial_\psi)\,, \qquad 
	J^3=-\frac{i}{2}(\partial_{\phi}+\partial_{\psi})\,,
	\\
	\tilde{J}^{\pm}&=\frac{1}{2}  e^{\pm i(\phi-\psi)} (\mp \partial_\theta -i \cot\theta  \partial_\phi- i \tan\theta \partial_\psi)
	\,, \qquad
	\tilde{J}^3=-\frac{i}{2}(\partial_{\phi}-\partial_{\psi})\,.
\end{split}
\end{equation}

One can generate the degree $k$ scalar spherical harmonics acting with the lowering operators in \eqref{generators isometry S3} on the highest spin, degree $k$ scalar spherical harmonics, which are 
\begin{equation}\label{highest sph harmonic} \begin{split}
Y_k^{\pm\frac{k}{2},\pm\frac{k}{2}}=\sqrt{k+1}\sin^k\theta e^{\pm i k \phi} \, .
\end{split}
\end{equation}
We make use of the degree $k=1,2$ normalized scalar spherical harmonics, given by:
\begin{equation}	\label{sph har k=1} \begin{split}
Y_1^{+\frac{1}{2},+\frac{1}{2}}= \sqrt{2} \sin\theta \,e^{i \phi}&\,, \qquad 
Y_1^{+\frac{1}{2},-\frac{1}{2}}=  \sqrt{2} \cos\theta \,e^{i \psi} \,,\\
Y_1^{-\frac{1}{2},+\frac{1}{2}}= - \sqrt{2} \cos\theta \,e^{-i \psi} &\,, \qquad 
Y_1^{-\frac{1}{2},-\frac{1}{2}}=  \sqrt{2} \sin\theta \,e^{-i \phi}\,;
\end{split}
\end{equation}

\begin{equation}\label{sph har k=2}	 \begin{split}
Y_2^{+1,+1}&=\sqrt{3} \sin^2\theta\, e^{2 i \phi}\,, \qquad
Y_2^{+1,0}= \sqrt{6} \sin\theta \cos\theta \,e^{ i (\phi+\psi)} \,, \qquad  Y_2^{+1,-1}=\sqrt{3} \cos^2\theta e^{2 i \psi} \,,\\
Y_2^{0,+1}&=-\sqrt{6} \sin\theta \cos\theta \,e^{ i (\phi-\psi)} \,, \qquad  Y_2^{0,0}=-\sqrt{3} \cos2\theta\,, \qquad  Y_2^{0,-1}=\sqrt{6} \sin\theta \cos\theta \,e^{ -i (\phi-\psi)} \,, \\
Y_2^{-1,+1}&=\sqrt{3} \cos^2\theta \,e^{-2 i \psi} \,, \qquad Y_2^{-1,0}=-\sqrt{6} \sin\theta \cos\theta \,e^{ -i (\phi+\psi)} \,, \qquad 
Y_2^{-1,-1}=\sqrt{3} \sin^2\theta \,e^{-2 i \phi} \,.
\end{split}
\end{equation}
We also introduce degree 1 vector spherical harmonics $Y^{a\pm}_1$ ($a=\pm,0$),
\begin{equation}\label{vector harmonics}
\begin{split}
Y_1^{++}&=\frac{1}{\sqrt{2}}e^{i(\phi+\psi)}\,\left[-i \,d\theta+\sin\theta\cos\theta\,d(\phi-\psi)\right]\,,\\
Y_1^{-+}&=\frac{1}{\sqrt{2}}e^{-i(\phi+\psi)}\,\left[ i \,d\theta+\sin\theta\cos\theta\,d(\phi-\psi)\right]\,,\\
Y_1^{0+}&=-\cos^2\theta\,d\psi-\sin^2\theta\,d\phi\,,\\
Y_1^{+-}&=\frac{1}{\sqrt{2}}e^{i(\phi-\psi)}\left[i \,d\theta-\sin\theta\cos\theta\, d(\phi+\psi)\right]\,,\\
Y_1^{--}&=-\frac{1}{\sqrt{2}}e^{-i(\phi-\psi)}\left[i \,d\theta+\sin\theta\cos\theta\, d(\phi+\psi)\right]\,,\\
Y_1^{0-}&=\cos^2\theta\,d\psi-\sin^2\theta\,d\phi\,,\\
\end{split}
\end{equation}
which are normalized as
\begin{equation}\label{normalization vec harmonic} 
\begin{split}
\int (Y_1^{a A})_i^* (Y_1^{b B})^{i} \;=\;\Omega_3\, \delta^{a,b} \delta^{A,B}\,,
\end{split}
\end{equation}
where $a,b=\pm,0$, $A,B=\pm$ and the index $i$ is raised and lowered with the metric on S$^3$.

We define the following triple integral:
\begin{equation}\label{triple overlap} \begin{split}
\int Y_k^{(m_k,\tilde{m_k})}\big(Y_1^{a-}\big)_i \big(Y_1^{b+}\big)^i\;=\;\Omega_3 f^{(k)}_{(m_k,\tilde{m_k}) a b}\,.
\end{split}
\end{equation}
The explicit value of the components of $f^{(k)}_{(m_k,\tilde{m_k}) ab}$, defined in~\eqref{triple overlap}, that have been used in this paper are
\begin{equation}\label{f overlap} \begin{split}
f^{(2)}_{(0,0)00}\,=\,\frac{1}{\sqrt{3}} \,,\hspace{2em}
f^{(2)}_{(1,1)--}\,=\,\frac{1}{\sqrt{3}} \,,\hspace{2em}
f^{(2)}_{(\pm1,\pm1)00}\,=\,0\,.
\end{split}
\end{equation}

\section{Computations of CFT correlators}
\label{sec:twistfieldcorr}

In this appendix we compute the CFT correlators \eq{correlator R sector ++1J-00k} and \eq{eq:corr-app-2}, which are used in Eqs.\;\eq{eq:use-app-2} and \eq{eq:use-app-1} respectively, using the method developed in \cite{Lunin:2000yv,Lunin:2001pw}. 
In order to do so, and for reference in the main part of the paper, we also record some conventions and notation.

\subsection{Conventions and notation}

On the CFT cylinder with coordinate $w=\tau+i\sigma$, the bare twist operator $\sigma_k$ corresponding to the permutation $(12\cdots k)$ is defined to introduce the following boundary conditions on the fields $X^i_{(r)}$, $\psi^{\alpha A}_{(r)}$, $r=1,2,\ldots k$ as they circle the insertion point $w_*$ (see e.g.~\cite[Eq.\;(2.12)]{Carson:2014ena}):
\bea
&&X_{(1)} \,\rightarrow\, X_{(2)} \,\rightarrow\, \cdots \,\rightarrow\, X_{(k)} \,\rightarrow\, X_{(1)} \,, \label{eq:bos-tw} \cr
&&\psi_{(1)} \,\rightarrow\, \psi_{(2)} \,\rightarrow\, \cdots \,\rightarrow\, \psi_{(k)} \,\rightarrow\, -\psi_{(1)} \,,
\eea
and likewise for the right-moving fermions. We have of course suppressed some indices to lighten the notation here.

In the full symmetric orbifold CFT, the bare twist operator $\Sigma_k$ is defined by symmetrizing $\sigma_k$ over all $k$-cycles:
\begin{equation}
 \Sigma_k \;=\; \sum_{k\mhyphen \text{cycles}}\sigma_k \,.
 \end{equation} 
The conformal dimension of $\sigma_k$ is $h=\bar{h}=\frac{1}{4}(k-\frac{1}{k})$ and it is neutral under $SU(2)_L\times SU(2)_R$. 

The insertion of a twist operator $\sigma_k$ allows the existence of fractional modes of the operators. Switching to the CFT plane with coordinate $z=e^w$, for a primary operator $O$ with conformal dimension $h$, these are defined by \cite{Lunin:2001pw}:
\begin{equation}\label{definition fractional modes}
O_{-\frac{m}{k}}\equiv \int \frac{dz}{2\pi i} \sum_{r=1}^k O_{(r)}(z)e^{-2\pi i \frac{m}{k} (r-1)}z^{-\frac{m}{k}+h-1}\,.
\end{equation}  
To construct a chiral primary operator starting from the bare twist $\sigma_k$ we must raise its charge until $h=j$ and $\bar{h}=\bar{j}$. This is achieved by using fractional modes of the current operators and, for even $k$, spin fields $S^\pm\,\,,\bar{S}^\pm$~\cite{Lunin:2001pw}. A set of twist-$k$ chiral primary operators is given by~\cite{Lunin:2001pw} (see also~\cite{Avery:2010qw,Carson:2014ena})
\begin{equation}\label{chiral primary twist operator}
\sigma^{\frac{k-1}{2},\frac{k-1}{2}}_k \;\equiv\;
     \begin{cases}
    \tilde{J}^+_{-(k-2)/k}...\tilde{J}^+_{-1/k} J^+_{-(k-2)/k}...J^+_{-1/k}\sigma_k(z) \hspace{5em} k \text{ odd}\\
     \tilde{J}^+_{-(k-2)/k}...\tilde{J}^+_{-2/k}J^+_{-(k-2)/k}...J^+_{-2/k}(S^+\bar{S}^+)\sigma_k(z) \hspace{2em} k \text{ even}\\
     \end{cases}
\end{equation}
These operators have dimension and charge $h=\bar{h}=j=\bar{j}=\frac{k-1}{2}$.

The covering-space method of~\cite{Lunin:2000yv,Lunin:2001pw} for computing correlators of twist operators involves mapping to a local covering space (with coordinate $t$), given by a map that is locally of the form
\begin{equation}\label{map covering space}
z-z_* \;\simeq\; b_* (t-t_*)^k \;.
\end{equation}
The $k$ sets of fields $X^i_{(r)}$, $\psi^{\alpha A}_{(r)}$, which had untwisted boundary conditions in the absence of the twist operator, are mapped to one set of single-valued fields in the covering space, and in the $t$-plane there are no twist operator insertions; the only parts of the $z$-plane operator \eqref{chiral primary twist operator} that survive in the covering $t$-plane  are the currents and spin fields. When the operator is inserted at the origin of the $t$-plane, the spin fields in \eq{chiral primary twist operator} for even $n$ create the RR vacuum $\ket{++}^{(t)}\,$. For more discussion and recent related work, see e.g.~\cite{Carson:2014ena,Carson:2014yxa,Carson:2014xwa,Avery:2010er,Burrington:2012yn,Burrington:2012yq,Burrington:2015mfa,Burrington:2017jhh,Burrington:2018upk,deBeer:2019ioe,Tormo:2018fnt}.

Passing to the $t$-plane via the map~\eqref{map covering space}, one obtains the following relation between the modes in the $z$-plane (given in Eq.\;\eqref{definition fractional modes}) and those in the covering space:
\begin{equation}\label{relation modes z t plane}
O_{-\frac{m}{k}}^{(z)}~\rightarrow~\int\frac{dt}{2\pi i}  \left(\frac{dz}{dt}\right)^{\!\!-h+1}\!\! O(t) \;\! \big(b_t t^k\big)^{-\frac{m}{k}+h-1}~=~b_t^{-\frac{m}{k}}k^{1-h}O^{(t)}_{-m}
\end{equation}
where the superscripts $(z)$ and $(t)$ distinguish the operators in the $z$-plane from those in the $t$-plane.

We are interested in (normalized) three-point functions in the $z$-plane of the following form:
\begin{equation}\label{correlator O1O2O3}
\frac{\langle O_1^\dagger(\infty)O_2(a)O_3(0)\rangle}{\langle O_1^\dagger(\infty)O_1(0)\rangle}
\end{equation}
where $O_i$ is an operator that is composed of a bare twist contribution $\sigma_{k_i}$ with conformal dimension $h_{i}=\frac{1}{4}(k_i-\frac{1}{k_i})$ and a spin contribution which we denote schematically by $S_i$, i.e. $O_i\,=\,S_i\sigma_{k_i}\,$. As discussed in~\cite[Eq.\;(3.18)]{Lunin:2001pw}, the contributions of the twist fields and the spin fields in the correlator~\eqref{correlator O1O2O3} factorize as follows:
\begin{equation}\label{correlator split twist spin}
\frac{\langle O_1^\dagger(\infty)O_2(a)O_3(0)\rangle}{\langle O_1^\dagger(\infty)O_1(0)\rangle}\;=\;|C_{1,2,3}|^{12}|a|^{-2(h_{1}+h_{2}-h_{3})}\frac{\langle S_1^\dagger(\infty)S_2(a)S_3(0)\rangle}{\langle S_1^\dagger(\infty)S_1(0)\rangle}
\end{equation}
where the $a$ dependence is given by conformal invariance and $C_{1,2,3}$ is the fusion coefficient of a bosonic theory with $c=1$. The exponent 12 appears because we have $c=6$ on a single copy.  The coefficient $C_{1,2,3}$ was computed in~\cite{Lunin:2000yv}; we will thus focus on the spin field correlator, which can be computed using bosonization~\cite{Lunin:2001pw}.

We introduce holomorphic (antiholomorphic) bosonic fields $\phi_5(z)$ and $\phi_6(z)$ ($\tilde{\phi}_5(\bar{z})$ and $\tilde{\phi}_6(\bar{z})$). We bosonize the fermions as (for brevity we write only the holomorphic expressions)
\begin{equation}\label{definition bosonization-2}
\psi^{++}\,=\,e^{i\phi_5} \,, \qquad \psi^{+-} \,=\, e^{-i\phi_6}\,, \qquad
\psi^{-+}\,=\,e^{i\phi_6}\,, \qquad  \psi^{--}\,=\,- e^{-i\phi_5} \,.
\end{equation}
Normal ordering is implicit as usual, and we shall suppress cocycles as these will not be important for our purposes.
We also introduce the notation 
\be
\qquad\qquad  \phi_{-} \, \equiv\,\phi_5 -\phi_6  \qquad\quad \Rightarrow \quad 
e^{i \alpha \phi_-(z)}e^{i \beta \phi_-(w)} ~\sim~  e^{i \alpha \phi_-(z)+i \beta \phi_-(w)} (z-w)^{2\alpha\beta} \,.
\ee
In terms of $\phi_{-}$, the $SU(2)_L$ currents are
\begin{equation}\label{bosonized currents-2}
J^+(z)\,=\,e^{i \phi_-(z)}\,, \qquad~~ J^-(z)\,=\,e^{-i\phi_-(z)}\,, \qquad J^3(z)\,=\,\frac{i}{2}\partial\phi_- (z) \,.
\end{equation}
We will also need the expression for the operator $O^{--}$:
\begin{equation}\label{bosonized O--}
O^{--}(z,\bar{z})\;=\;\frac{1}{\sqrt{2}}\!\!\:\left(e^{-i\phi_5(z)+i\tilde{\phi}_6(\bar{z})}-e^{i{\phi}_6({z})-i\tilde{\phi}_5(\bar{z})}\right).
\end{equation}
Using Eqs.\;\eqref{relation modes z t plane} and~\eqref{bosonized currents-2} we obtain the following expression for the twist-$k$ primaries in Eq.\;\eqref{chiral primary twist operator} lifted to the covering space:
\begin{equation} \label{eq:cp-cover}
\qquad \sigma^{\frac{k-1}{2},\frac{k-1}{2}}_k(t,\bar{t}) \;=\; |b|^{-\frac{2p^2}{k}}e^{ip \:\! \phi_- (t)}e^{ip\:\!\tilde{\phi}_-(\bar{t})}\,, \qquad\qquad p\,=\,\tfrac{k-1}{2} \,.
\end{equation}
%

We conclude this subsection by recording our conventions for spectral flow. Spectral flow acts on states and operators as
\begin{equation}\label{spectral flow transformation}
\ket{\phi}\,\rightarrow\,\ket{\phi'}\,=\,U_\nu \ket{\phi} \,, \hspace{2em}
O\,\rightarrow O'\,=\,U_\nu \:\! O \:\! U_\nu^\dagger \;,
\end{equation}
where $U_\nu=e^{i \nu \phi_-}$ in the $z$-plane, and where on the covering $t$-plane of a strand of length $k$, $U_\nu=e^{i k \nu \phi_-}$. Analogous expressions hold in the antiholomorphic sector.
The spectral flow transformations for the modes of the $SU(2)_L$ currents are (for the rest of the chiral algebra, see e.g.~\cite[App.\;A]{Avery:2010qw}):\footnote{The convention map to the spectral flow parameters of \cite{Avery:2010qw} is $\alpha_{\rm there}=2\nu_{\rm here}$.}
\begin{equation} \label{eq:sf-J}
J^3_m\,\rightarrow\, J^3_m - \frac{c \:\! \nu }{6}\delta_{m,0} \,, \hspace{4em} J^\pm_m\,\rightarrow\, J^\pm_{m\mp 2\nu} \,.
\end{equation}
The weight and $SU(2)$ charge $(h,j)$ of states transform as
\be \label{eq:sf-states}
h \;\to\; h + 2 \nu j + \frac{c \:\! \nu^2}{6} \,, \qquad\quad j\;\to\; j+\frac{c\:\! \nu}{6} \,;
\ee
for example, spectral flow with parameters $(\nu,\bar\nu)=(\frac12,\frac12)$ maps the NS-NS vacuum to the RR ground state $\ket{++}$.

\subsection{Expectation value of $\Sigma_3^{00}$ on a 3-charge state}
\label{sec:Sigma3app}
We are now ready to compute the following normalized three-point function, for use in Eqs.\;\eq{eq:use-app-2}: 
\begin{equation}\label{correlator R sector ++1J-00k}
\frac{ {}_1\bra{++}{}_k\bra{00}J^{-}_{+1}\:\! \sigma_3^{00}(a)\:\! J^+_{-1}\ket{00}_k\ket{++}_1}
{{}_1\bra{++}{}_k\bra{00}J^{-}_{+1}J^+_{-1}\ket{00}_k\ket{++}_1}
~\equiv~ C_{k3k}^{00(m=1)} |a|^{-2} \,.
\end{equation}
In order to exploit the machinery worked out so far, we map the correlator Eq.\;\eqref{correlator R sector ++1J-00k} to the NS-NS sector using spectral flow with parameters $(-\frac12,-\frac12)$. This is a unitary transformation \eq{spectral flow transformation} that leaves invariant the value of the correlator. 

This spectral flow transformation maps the RR vacuum $\ket{++}_1$ to the untwisted NS-NS vacuum. From \eq{eq:sf-J}, the operator $J^+_{-1}$ becomes $J^+_0$ in the NS sector.

Next, $\sigma_3^{00}$ is defined by  $\sigma_3^{00}\equiv \frac{1}{2}[J^-_0,[\tilde{J}_0^-,\sigma_3^{11}]]$. Using Eqs.\;\eqref{chiral primary twist operator} and \eqref{eq:cp-cover}, we see that in the covering space  $\sigma_3^{00}\to 2\:\!|b|^{-2/3} J^3\tilde{J}^3$, so this operator is invariant under spectral flow.

To derive the spectral flow of the state $\ket{00}_k$, recall that it is defined by:
\begin{equation} \label{eq:00}
\ket{00}_k\;\equiv\;O^{--}_0\ket{++}_k\;=\;\frac{1}{\sqrt{2}}\epsilon_{\dot{A}\dot{B}}\psi_0^{-\dot{A}}\tilde{\psi}_0^{-\dot{B}}\ket{++}_k \,.
 \end{equation} 
In the $z$-plane, $\psi^{-\dot{A}}_{0}$ is spectral flowed to $\psi^{-\dot{A}}_{-\frac12}$, which is related to the corresponding covering-space mode through Eq.\;\eqref{relation modes z t plane}, giving
\begin{equation}
\begin{split}
\psi_{-\frac12}^{-\dot{A}}~\rightarrow ~ b_t^{-\frac12} \!\!\: \sqrt{k} \;\!  \psi_{-\frac{k}{2}}^{-\dot{A}(t)} \;.
\end{split}
\end{equation}
Under spectral flow with parameters $(-\frac12,-\frac12)$, the RR ground state $\ket{++}_k$ is mapped to an anti-chiral primary state;  moving to the covering $t$-plane gives (as usual normal ordering of exponentials should be understood; we leave this implicit to lighten the notation)
\begin{equation}
\ket{++}_k ~\rightarrow ~
\sigma_k^{-\frac{k-1}{2},-\frac{k-1}{2}}\ket{0}_{\NS}^{(t)} ~=~ |b|^{-\frac{2p^2}{k}}e^{-i p\:\!\phi_-(0)}e^{-i p\:\!\tilde{\phi}_-(0)}\ket{0}_{\NS}^{(t)} 
\end{equation}
where $\ket{0}_{\NS}^{(t)}$ is the NS-NS vacuum of the covering $t$-plane and again $p=\frac{k-1}{2}$.
Then from Eqs.\;\eq{bosonized O--} and \eq{eq:00} we obtain 
\bea 
\!\!\!\!\!\!\!\!\!\!\!\!&&\ket{00}_k  ~\rightarrow ~
\frac{\sqrt{k}}{\sqrt{2}}|b|^{-\frac{2p^2}{k}-1}\Big(\psi^{--(t)}_{-\frac{k}{2}}\tilde{\psi}^{-+(t)}_{-\frac{k}{2}}-\psi^{-+(t)}_{-\frac{k}{2}}\tilde{\psi}^{--(t)}_{-\frac{k}{2}}\Big)e^{-ip\:\!\phi_-(0)}e^{-ip\:\!\tilde{\phi}_-(0)}\ket{0}_{\NS}^{(t)}
\cr
\!\!\!\!\!\!\!\!\!\!\!\!&&=
\frac{\sqrt{k}}{\sqrt{2}}|b|^{-\frac{2p^2}{k}-1}\!\int\!\!\!\;\frac{dt d\bar{t}}{(2\pi i)^2}t^{-\frac{k+1}{2}}\bar{t}^{-\frac{k+1}{2}}\big(e^{-i\phi_5(t)+i\tilde{\phi}_6(\bar{t})}-e^{i{\phi}_6({t})-i\tilde{\phi}_5(\bar{t})}\big)e^{-ip\:\!\phi_-(0)}e^{-ip\:\!\tilde{\phi}_-(0)}\ket{0}_{\NS}^{(t)}
\cr
\!\!\!\!\!\!\!\!\!\!\!\!&&=
\frac{\sqrt{k}}{\sqrt{2}}|b|^{-\frac{2p^2}{k}-1}\!\int\!\!\!\;\frac{dt d\bar{t}}{(2\pi i)^2}t^{-1}\bar{t}^{-1}\Big(e^{-i\phi_5(t)-ip\:\!\phi_-(0)+i\tilde{\phi}_6(\bar{t})-ip\:\!\tilde{\phi}_-(0)}
-e^{i\phi_6(t)-ip\:\!\phi_-(0)-i\tilde{\phi}_5(\bar{t})-ip\:\!\tilde{\phi}_-(0)}\Big)\ket{0}_{\NS}^{(t)}
\cr
\!\!\!\!\!\!\!\!\!\!\!\!&&=
\frac{\sqrt{k}}{\sqrt{2}}|b|^{-\frac{2p^2}{k}-1}\bigg(e^{-i\frac{k+1}{2}\phi_5(0)+i\frac{k-1}{2} \phi_6(0)-i\frac{k-1}{2}\tilde{\phi}_5(0)+i\frac{k+1}{2} \tilde{\phi}_6(0)}  \cr
\!\!\!\!\!\!\!\!\!\!\!\!&&\hspace{6cm}
 - e^{-i\frac{k-1}{2}\phi_5(0)+i\frac{k+1}{2} \phi_6(0)-i\frac{k+1}{2} \tilde{\phi}_5(0)+i\frac{k-1}{2} \tilde{\phi}_6(0)}\bigg)
\ket{0}_{\NS}^{(t)}
\eea
so we obtain 
\begin{equation}\label{J+-100k++1 ket NS cover}
\begin{split}
&J^+_{-1}\ket{00}_k ~\rightarrow~ \frac{\sqrt{k}}{\sqrt{2}}|b|^{-\frac{2p^2}{k}-1}\bigg(e^{-i\frac{k-1}{2}\phi_5(0)+i\frac{k-3}{2} \phi_6(0)-i\frac{k-1}{2}\tilde{\phi}_5(0)+i\frac{k+1}{2} \tilde{\phi}_6(0)} \\
&\hspace{6cm} 
-e^{-i\frac{k-3}{2}\phi_5(0)+i\frac{k-1}{2} \phi_6(0)-i\frac{k+1}{2} \tilde{\phi}_5(0)+i\frac{k-1}{2} \tilde{\phi}_6(0)}\bigg)
\ket{0}_{\NS}^{(t)}  \,.
\end{split}
\end{equation}
Before computing the spin correlator in Eq.\;\eqref{correlator split twist spin}, we recall the value of the twist fusion coefficient we require \cite[Eq.\;(6.25)]{Lunin:2000yv},
\begin{equation}\label{bare twish contribution 3kk}
|C_{k,3,k}|^{12}\;=\;\frac{(k+1)^{\frac{k^2+1}{k}+\frac23}}{2^{\frac{4}{3}}3^{\frac53}k^{\frac43}(k-1)^{\frac{k^2+1}{k}-\frac23}} \,.
\end{equation}
The map from the $z$-plane to the covering space used to compute this is given in \cite[Eq.\;(4.34)]{Lunin:2000yv}. Since it will be needed in the following, we report here the behaviour of this map near the insertion points $z=0,a,\infty$, as given in \cite[Eq.\;(6.18)--(6.20)]{Lunin:2001pw}\footnote{We note a typo in \cite[Eq.\;(6.20)]{Lunin:2001pw}: $(d-1-d_2) \to (d_1-d_2)$.}:
\begin{equation}\label{asymp map covering space}
\begin{split} 
&z\,\sim\,  b_0t^k\,=\,a\frac{k+1}{k-1}t^k \hspace{14.9em} \text{near }z=0 \,, \\
&z\,\sim\, a+b_1(t-1)^3\,=\, a+a\frac{(k+1)k(k-1)}{12}(t-1)^3 \hspace{2em} \text{near }z=a \,, \\
&z\,\sim\,  b_\infty t^k\,=\,a\frac{k-1}{k+1} t^k \hspace{14.5em} \text{near }z=\infty \,.
\end{split}
\end{equation}
Note that the map has been chosen so that the point $z=a$ is mapped to $t=1$.
Because of the normalization in~\eqref{correlator split twist spin}, the spin field correlator in the case of~\eqref{correlator R sector ++1J-00k} arises from contracting the covering-space operator $\sigma_3^{00(t)}(1)\,=\, 2|b_1|^{-\frac23}J^3\tilde{J}^3(1)\,=\,-\frac{1}{2}|b_1|^{-\frac23}\partial \phi_- \bar\partial\tilde{\phi}_-(1)$ with the operator in Eq.\;\eqref{J+-100k++1 ket NS cover}. The result reads:
\begin{equation}
\qquad
\frac{\langle S_1^\dagger(\infty)S_2(a)S_3(0)\rangle}{\langle S_1^\dagger(\infty)S_1(0)\rangle}\;=\;
\frac{k(k-2)}{2}|b_0|^{-\frac{2p^2}{k}-1}|b_1|^{-\frac{2}{3}}|b_{\infty}|^{\frac{2p^2}{k}+1}
\,, \qquad\quad~~ p\,=\,\tfrac{k-1}{2} \,.
\end{equation}
Using Eqs.\;\eqref{correlator split twist spin},~\eqref{bare twish contribution 3kk} and~\eqref{asymp map covering space} we find
\begin{equation}
 C_{k3k}^{00(m=1)}\;=\;\frac{k-2}{6k} \,.
\end{equation} 

\subsection{Expectation value of $\Sigma_2^{++}$ in the state $\ket{00}_1\ket{++}_2$}

In Eq.\;\eq{eq:use-app-1} we make use of the relation $\sigma_2^{++}\ket{00}_1\ket{00}_1=-\frac{1}{4}\ket{++}_2$, which we now derive. The coefficient corresponds to computing the following correlator:
\begin{equation} \label{eq:corr-app-2}
\frac{{}_2\bra{++}\sigma_2^{++}\ket{00}_1\ket{00}_1}{{}_2\bra{++}\ket{++}_2} \,.
\end{equation}
We lift this correlator to the covering space with the map (c.f.~\cite{Carson:2014ena})
\begin{equation}\label{covering map 001++2}
z\,=\,t(t-1)\,.
\end{equation}
The point $z=0$ corresponds to the points $t=0,1$, where we have the insertions
\begin{equation}
(O^{--}S^+\bar{S}^+)(0)\,\,,\quad (O^{--}S^+\bar{S}^+)(1) \,.
\end{equation}
Writing only the holomorphic expressions, the operators $S^\pm$  take the following form in the covering space:
\begin{equation}
S^\pm(t)\;=\;|b_t|^{-\frac{1}{4k}}e^{\pm\frac{i}{2}\phi_-(t)} \,.
\end{equation}
The operator $\sigma_2^{++}$ is inserted at the point $t=1/2$ in the covering space (see e.g.~\cite[Sec.\;4]{Carson:2014ena}).
The asymptotic behaviour of the map~\eqref{covering map 001++2} at the insertion points $t=0,1/2,1,\infty$ is given in~\cite[Eq.\;(C.45)]{Carson:2014ena}. 
Factorizing the correlator as in~\eqref{correlator split twist spin}, we compute the spin contribution
\begin{equation}\label{spincorrelator 001++2}
\frac{\langle(S^-\bar{S}^-)(\infty)(S^+\bar{S}^+)(\frac{1}{2}) (O^{--}S^+\bar{S}^+)(1)(O^{--}S^+\bar{S}^+)(0) \rangle}{\langle(S^-\bar{S}^-)(\infty)(S^+\bar{S}^+)(0)\rangle}~=~- |b_\infty|^{\frac{1}{4}}|b_{\frac{1}{2}}|^{-\frac{1}{4}}|b_1|^{-\frac{1}{2}}|b_0|^{-\frac{1}{2}}\,.
\end{equation}
The remaining contributions to the correlator are given in~\cite[Eqs.\;(C.5),\;(C.39)]{Carson:2014ena}. Combining these results with~\eqref{spincorrelator 001++2}, we obtain the value of the desired coefficient,
\begin{equation}
\label{eq:app-b-result}
\sigma_2^{++}\ket{00}_1\ket{00}_1\,=\,-\frac{1}{4}\ket{++}_2 \,.
\end{equation}

\newpage
\section{Type IIB supergravity ansatz and BPS equations}\label{app:sugra}

The general solution to Type IIB supergravity compactified on T$^4$ that is 1/8-BPS, has D1-D5-P charges, and is invariant on the T$^4$ directions is~\cite[Appendix E.7]{Giusto:2013rxa}:
\begin{subequations}\label{ansatzSummary}
\allowdisplaybreaks
 \begin{align}
d s^2_{10} &~=~ \sqrt{\alpha} \,ds^2_6 +\sqrt{\frac{Z_1}{Z_2}}\,d \hat{s}^2_{4}\, ,\label{10dmetric}\\
d s^2_{6} &~=~-\frac{2}{\sqrt{\cP}}\,(d v+\betab)\,\Big[d u+\omega + \frac{\mathcal{F}}{2}(d v+\betab)\Big]+\sqrt{\cP}\,d s^2_4\,,\\
e^{2\Phi}&~=~\frac{Z_1^2}{\cP}\, ,\\
B&~=~ -\frac{Z_4}{\cP}\,(d u+\omega) \wedge(d v+\betab)+ a_4 \wedge  (d v+\betab) + \gamma_4\,, \label{Bform}\\ 
C_0&~=~\frac{Z_4}{Z_1}\, ,\\
C_2 &~=~ -\frac{Z_2}{\cP}\,(d u+\omega) \wedge(d v+\betab)+ a^1 \wedge  (d v+\betab) + \gamma_2\,,\\ 
C_4 &~=~ \frac{Z_4}{Z_2}\, \widehat{\mathrm{vol}}_{4} - \frac{Z_4}{\cP}\,\gamma_2\wedge (d u+\omega) \wedge(d v+\betab)+x_3\wedge(d v + \betab) \,, \\
C_6 &~=~\widehat{\mathrm{vol}}_{4} \wedge \left[ -\frac{Z_1}{\cP}\,(d u+\omega) \wedge(d v+\betab)+ a^2 \wedge  (d v+\betab) + \gamma_1\right] , 
\end{align}
\end{subequations}
where
\begin{equation}
\alpha \;=\; \frac{Z_1 Z_2}{Z_1 Z_2 - Z_4^2}~,\qquad~~
\cP   \;=\;     Z_1  Z_2  -  Z_4^2 \,.
\label{Psimp}
\end{equation}
In the above, $d \hat{s}^2_4$ denotes the flat metric on $T^4$, and $\widehat{\mathrm{vol}}_{4}$ stands for the corresponding volume form. This ansatz contains all fields known to arise from worldsheet calculations of the backreaction of D1-D5-P bound states invariant on $\cM$~\cite{Giusto:2011fy}.

The BPS equations have the following structure. The base metric, $ds^2_4$, and the one-form $\betab$ satisfy non-linear equations. Having solved these initial equations, the remaining ansatz quantities are organized into two layers of linear equations~\cite{Giusto:2013rxa,Bena:2011dd}.

We denote the exterior differential on the spatial base $\cB$ by $\tilde d$, and introduce~\cite{Gutowski:2003rg} 
\begin{equation}
\mathcal{D} \;\equiv\; \tilde d - \betab\wedge \frac{\partial}{\partial v}\,.
\end{equation}

In the present paper we consider only solutions where the four-dimensional base space is flat $\mathbb{R}^4$, and in which $\betab$ does not depend on $v$.
Then the BPS equation for $\betab$ is 
 \begin{equation}\label{eqbeta}
d \betab \;=\; *_4 d\betab\,,
 \end{equation}
where $*_4$ denotes the flat $\mathbb{R}^4$ Hodge dual.

To write the remaining BPS equations in a covariant form, we rescale $(Z_4,a_4,\gamma_4) \to (Z_4,a_4,\gamma_4)/\sqrt{2}$ for the remainder of this appendix (and only here).  We introduce the $SO(1,2)$ Minkowski metric $\eta_{ab}$ ($a=1,2,4$) via
\be \label{eq:etaab}
\eta_{12} ~=~ \eta_{21} ~=~ 1\,, \qquad \eta_{44} = -1 \,.
\ee
This is used to raise and lower $a,b$ indices. 
We introduce the two-forms $\Theta^1$, $\Theta^2$, $\Theta^4$ via\footnote{The relation to the notation of~\cite{Bena:2017xbt} is that $\Theta^1_{\mathrm{here}}=\Theta_1^{\mathrm{there}}$, $\Theta^2_{\mathrm{here}}=\Theta_2^{\mathrm{there}}$, $(1/\sqrt{2})\Theta_4^{\mathrm{here}}=\Theta_4^{\mathrm{there}}$.}
\begin{equation}
\label{Thetadefs}
\Theta^b ~\equiv~ \mathcal{D} a^b + \eta^{bc} \:\! \dot{\gamma}_c  \;.
\end{equation}
We now have
\be
\cP ~\equiv~ \coeff{1}{2} \eta^{ab} Z_a Z_b ~=~ Z_1 Z_2 - \coeff12 Z_4^2 \,.
\ee
The first layer of the BPS equations then takes the form
\bea \label{eq:firstlayer}  
 *_4 D\dot{Z}_a ~=~ & \eta_{ab} D\Theta^{b}\,,\qquad D*_4 DZ_a ~=~  - \eta_{ab} \Theta^{b} \! \wedge d\betab\,,
\qquad \Theta^{a} ~=~ *_4 \Theta^{a} \,.
\eea
The second layer becomes
\begin{equation}
 \begin{aligned}
D \omega + *_4 D\omega + \mathcal{F} \,d\betab 
~=~ & Z_a \Theta^{a}\,,  \\ 
 *_4D*_4\!\Bigl(\dot{\omega} -\coeff{1}{2}\,D\mathcal{F}\Bigr) 
~=~& \ddot \cP  -\coeff{1}{2} \eta^{ab} \dot{Z}_a \dot{Z}_b 
-\coeff{1}{4} \eta_{ab} *_4\! \Theta^{a}\wedge \Theta^{b} \,.
\end{aligned}
\label{eqFomega-app}
\end{equation} 
%


\newpage

\begin{adjustwidth}{-3mm}{-3mm} 
\bibliographystyle{utphys}      
\bibliography{microstates}       

\end{adjustwidth}


\end{document}